\documentclass[apj]{emulateapj}
\usepackage{color}

\slugcomment{accepted to The Astrophysical Journal}
\shorttitle{}
\shortauthors{Sano et al.}

\begin{document}
\title{Non-thermal X-rays and interstellar gas toward the $\gamma$-ray supernova remnant\\RX J1713.7$-$3946: Evidence for X-ray enhancement around CO and H{\sc i} clumps}

\author{H. Sano\altaffilmark{1}, T. Tanaka\altaffilmark{2}, K. Torii\altaffilmark{1}, T. Fukuda\altaffilmark{1}, S. Yoshiike\altaffilmark{1}, J. Sato\altaffilmark{1}, H. Horachi\altaffilmark{1}, T. Kuwahara\altaffilmark{1},\\T. Hayakawa\altaffilmark{1}, H. Matsumoto\altaffilmark{1}, T. Inoue\altaffilmark{3}, R. Yamazaki\altaffilmark{3}, S. Inutsuka\altaffilmark{1}, A. Kawamura\altaffilmark{1,5},\\K. Tachihara\altaffilmark{1}, H. Yamamoto\altaffilmark{1}, T. Okuda\altaffilmark{1}, N. Mizuno\altaffilmark{1,4}, T. Onishi\altaffilmark{1,5}, A. Mizuno\altaffilmark{6} and Y. Fukui\altaffilmark{1}}

\affil{$^1$Department of Physics, Nagoya University, Furo-cho, Chikusa-ku, Nagoya, 464-8601, Japan; sano@a.phys.nagoya-u.ac.jp}
\affil{$^2$Department of Physics, Kyoto University, Kitashirakawa-oiwake-cho, Sakyo-ku, Kyoto 606-8502, Japan}
\affil{$^3$Department of Physics and Mathematics, Aoyama Gakuin University, Fuchinobe, Chuou-ku, Sagamihara, 252-5258, Japan}
\affil{$^4$National Astronomical Observatory of Japan, Mitaka, 181-8588, Japan}
\affil{$^5$Department of Astrophysics, Graduate School of Science, Osaka Prefecture University, 1-1 Gakuen-cho, Naka-ku, Sakai, 599-8531, Japan}
\affil{$^6$Solar-Terrestrial Environment Laboratory, Nagoya University, Furo-cho, Chikusa-ku, Nagoya, 464-8601, Japan}

\begin{abstract}
RX J1713.7$-$3946 is the most remarkable very-high-energy $\gamma$-ray supernova remnant which emits synchrotron X-rays without thermal features. We made a comparative study of CO, H{\sc i} and X-rays in order to better understand the relationship between the X-rays, and the molecular and atomic gas. The results indicate that the X-rays are enhanced around the CO and H{\sc i} clumps on a pc scale but are decreased inside the clumps on a 0.1 pc scale. Magnetohydrodynamic numerical simulations of the shock interaction with molecular and atomic gas indicate that the interaction between the shock waves and the clumps excite turbulence which amplifies the magnetic field around the clumps \citep{inoue2012}. We suggest that the amplified magnetic field around the CO and H{\sc i} clumps enhances the synchrotron X-rays and possibly the acceleration of cosmic-ray electrons.
\end{abstract}


\section{Introduction}
RX J1713.7$-$3946 is one of the most prominent supernova remnants (SNRs) emitting high energy radiation covering the $\gamma$-rays and X-rays, which is likely emitted by cosmic-ray (CR) particles accelerated in the SNR via diffusive shock acceleration \citep[DSA; e.g.,][]{bell1978, blandford1978}. The SNR is located relatively close to the Galactic center at ($l$, $b$) = ($347$\fdg$3$, $-0$\fdg$5$), where contamination by the Galactic foreground/background is heavy at any wavelength. The SNR was not known as an SNR in the radio continuum emission and was discovered in the X-rays by the $ROSAT$ \citep{pfeffermann1996}. The VHE (very high energy) $\gamma$-rays from the SNR were discovered and mapped by the atmospheric Cherenkov telescopes \citep{enomoto2002, aharonian2004, aharonian2006a, aharonian2006b, aharonian2007}. In particular, the H.E.S.S. observations resolved that the $\gamma$-rays distribution is shell-like with a diameter of 1$^{\circ}$ by 0.1-degree diameter PSF (point spread function). Since the $\gamma$-rays are detected at an energy range 1--10 TeV, the CR protons producing the $\gamma$-rays may reach an energy range 10--800 TeV which is close to the knee, if the hadronic origin is working. The SNR is therefore an important candidate where the acceleration of the high energy CR protons is best tested in the Galaxy. It is noteworthy that the X-rays of the SNR are purely non-thermal synchrotron emission, indicating that the CR electrons are accelerated in the SNR up to the 10 TeV range; there are only two SNRs except for RX J1713.7$-$3946, which show such non-thermal X-rays, RX J0852.0$-$4622 (Vela Jr.) and HESS J1731$-$347, known to date \citep[e.g.,][]{koyama1997, slane2001, tian2010}. Detailed theoretical modeling of these high-energy radiation has been made over a wide range of physical parameters appropriate for the SNR, and has shown that the observed properties of high energy radiation are reproduced under reasonable sets of physical parameters relevant for the CR acceleration \citep[e.g.,][]{zirakashvili2007,zirakashvili2010}. It is thus becoming more and more important to constrain observationally the physical parameters of the SNR, like the magnetic field, and their distributions.

\begin{figure}
\begin{center}
\includegraphics[width=88mm,clip]{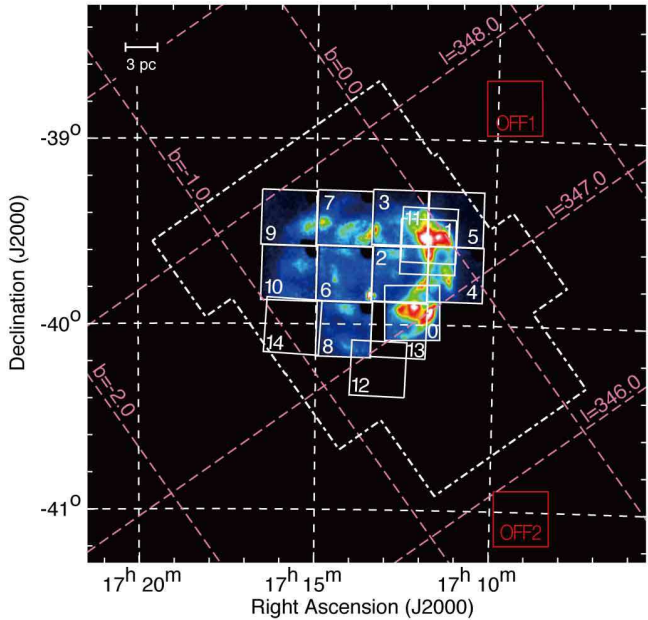}
\caption{$Suzaku$ FoV of each observation toward RX J1713.7$-$3946 overlaid on the $Suzaku$ XIS 0+2+3 mosaic image (1--5 keV) taken from \cite{tanaka2008}. The small squares correspond to the FoV of the XIS. The numbers indicated in the XIS FoV are pointing IDs used throughout this paper (see also Table \ref{tab1}). We also show the observed area in the $^{12}$CO($J$=2--1) enclosed by the white dash-dotted lines.}
\label{fig1}
\end{center}
\end{figure}%

It is now well established that the SNR is located at 1 kpc from us as first determined by the radial velocity of the associated CO molecular gas at $-7$ km s$^{-1}$ in $V_{\mathrm{LSR}}$ \citep{fukui2003}. This distance is confirmed by subsequent observations in X-rays and CO \citep{cassamchenai2004, moriguchi2005}. \cite{moriguchi2005} showed further details of the $^{12}$CO($J$=1--0) distribution and confirmed the interacting molecular gas identified by \cite{fukui2003}. At 1 kpc, it is most likely that the SNR corresponds to the historical SNR SN 393 with an age of 1600 yrs \citep{wang1997}. \citeauthor{sano2010} (2010, hereafter paper I) presented a comparison of the X-rays and the CO in $^{12}$CO($J$=2--1, 4--3) transitions for the western rim of the SNR, where most prominent mm/sub-mm CO clumps are distributed. These authors find that a dense CO clump named ``peak C'' is a site of low-mass star formation and is surrounded by bright X-rays at a pc scale, and that the CO clump shows an anti-correlation with the X-rays at a sub-pc scale. According to the magnetohydrodynamic numerical simulations \citep{inoue2012}, the SNR shock-cloud interaction excites turbulence in highly inhomogeneous interstellar medium (ISM), which leads to amplification of magnetic field around dense clumps. These observational and theoretical results suggest a picture that the magnetic field becomes stronger around dense clumps in the SNR, leading to bright X-rays at a pc scale around them, whereas the X-rays become weaker inside the CO clumps where the CR electrons cannot penetrate at a sub-pc scale. This suggests that the clumpy ISM may play a crucial role in producing the X-ray distribution by the shock-cloud interaction, even when the DSA is a fundamental mechanism to create CR electrons.

\begin{deluxetable}{cccclc}
\vspace*{-1cm}
\tablecaption{Summary of the $Suzaku$ archive data of RX J1713.7$-$3946}
\label{table1}
\tablehead{&&$\alpha_{\mathrm{J2000}}, \delta_{\mathrm{J2000}}$\hspace*{-0.2cm}&\hspace*{-0.5cm}XIS Exp.\hspace*{-0.7cm} &&
\\\multicolumn{1}{c}{ID}\hspace*{-0.05cm}&ObsID\hspace*{-0.2cm}& ($^{\mathrm{h}}$ $^{\mathrm{m}}$ $^{\mathrm{s}}$), ($^{\circ}$ $\arcmin$ $\arcsec$)\hspace*{-0.2cm}&(ks)&\multicolumn{1}{c}{Date}\hspace*{-0.2cm}&SCI\hspace*{-0.1cm}}
\startdata
\hspace*{-0.15cm}\scalebox{0.9}[1.0]{0}\hspace*{-0.2cm}&\scalebox{0.9}[1.0]{100026010}\hspace*{-0.2cm}&\scalebox{0.9}[1.0]{17 12 17.0, $-$39 56 11}\hspace*{-0.2cm}&\hspace*{-0.2cm}\scalebox{0.9}[1.0]{69}\hspace*{-0.2cm}&\hspace*{-0.2cm}\scalebox{0.9}[1.0]{2005 September  26}\hspace*{-0.2cm}&\scalebox{0.9}[1.0]{OFF}\hspace*{-0.2cm}\\
\hspace*{-0.15cm}\scalebox{0.9}[1.0]{1}\hspace*{-0.2cm}&\scalebox{0.9}[1.0]{501063010}\hspace*{-0.2cm}&\scalebox{0.9}[1.0]{17 11 51.5, $-$39 31 13}\hspace*{-0.2cm}&\hspace*{-0.2cm}\scalebox{0.9}[1.0]{18}\hspace*{-0.2cm}&\hspace*{-0.2cm}\scalebox{0.9}[1.0]{2006 September  11}\hspace*{-0.2cm}&\scalebox{0.9}[1.0]{OFF}\hspace*{-0.2cm}\\
\hspace*{-0.15cm}\scalebox{0.9}[1.0]{2}\hspace*{-0.2cm}&\scalebox{0.9}[1.0]{501064010}\hspace*{-0.2cm}&\scalebox{0.9}[1.0]{17 12 38.0, $-$39 40 14}\hspace*{-0.2cm}&\hspace*{-0.2cm}\scalebox{0.9}[1.0]{21}\hspace*{-0.2cm}&\hspace*{-0.2cm}\scalebox{0.9}[1.0]{2006 September  11}\hspace*{-0.2cm}&\scalebox{0.9}[1.0]{OFF}\hspace*{-0.2cm}\\
\hspace*{-0.15cm}\scalebox{0.9}[1.0]{3}\hspace*{-0.2cm}&\scalebox{0.9}[1.0]{501065010}\hspace*{-0.2cm}&\scalebox{0.9}[1.0]{17 12 38.2, $-$39 22 15}\hspace*{-0.2cm}&\hspace*{-0.2cm}\scalebox{0.9}[1.0]{22}\hspace*{-0.2cm}&\hspace*{-0.2cm}\scalebox{0.9}[1.0]{2006 September  11}\hspace*{-0.2cm}&\scalebox{0.9}[1.0]{OFF}\hspace*{-0.2cm}\\
\hspace*{-0.15cm}\scalebox{0.9}[1.0]{4}\hspace*{-0.2cm}&\scalebox{0.9}[1.0]{501066010}\hspace*{-0.2cm}&\scalebox{0.9}[1.0]{17 11 04.5, $-$39 40 10}\hspace*{-0.2cm}&\hspace*{-0.2cm}\scalebox{0.9}[1.0]{21}\hspace*{-0.2cm}&\hspace*{-0.2cm}\scalebox{0.9}[1.0]{2006 September  12}\hspace*{-0.2cm}&\scalebox{0.9}[1.0]{OFF}\hspace*{-0.2cm}\\
\hspace*{-0.15cm}\scalebox{0.9}[1.0]{5}\hspace*{-0.2cm}&\scalebox{0.9}[1.0]{501067010}\hspace*{-0.2cm}&\scalebox{0.9}[1.0]{17 11 05.1, $-$39 22 10}\hspace*{-0.2cm}&\hspace*{-0.2cm}\scalebox{0.9}[1.0]{21}\hspace*{-0.2cm}&\hspace*{-0.2cm}\scalebox{0.9}[1.0]{2006 September  12}\hspace*{-0.2cm}&\scalebox{0.9}[1.0]{OFF}\hspace*{-0.2cm}\\
\hspace*{-0.15cm}\scalebox{0.9}[1.0]{6}\hspace*{-0.2cm}&\scalebox{0.9}[1.0]{501068010}\hspace*{-0.2cm}&\scalebox{0.9}[1.0]{17 14 11.6, $-$39 40 14}\hspace*{-0.2cm}&\hspace*{-0.2cm}\scalebox{0.9}[1.0]{21}\hspace*{-0.2cm}&\hspace*{-0.2cm}\scalebox{0.9}[1.0]{2006 September  13}\hspace*{-0.2cm}&\scalebox{0.9}[1.0]{OFF}\hspace*{-0.2cm}\\
\hspace*{-0.15cm}\scalebox{0.9}[1.0]{7}\hspace*{-0.2cm}&\scalebox{0.9}[1.0]{501069010}\hspace*{-0.2cm}&\scalebox{0.9}[1.0]{17 14 11.4, $-$39 22 15}\hspace*{-0.2cm}&\hspace*{-0.2cm}\scalebox{0.9}[1.0]{18}\hspace*{-0.2cm}&\hspace*{-0.2cm}\scalebox{0.9}[1.0]{2006 September  19}\hspace*{-0.2cm}&\scalebox{0.9}[1.0]{OFF}\hspace*{-0.2cm}\\
\hspace*{-0.15cm}\scalebox{0.9}[1.0]{8}\hspace*{-0.2cm}&\scalebox{0.9}[1.0]{501070010}\hspace*{-0.2cm}&\scalebox{0.9}[1.0]{17 14 11.8, $-$39 58 14}\hspace*{-0.2cm}&\hspace*{-0.2cm}\scalebox{0.9}[1.0]{21}\hspace*{-0.2cm}&\hspace*{-0.2cm}\scalebox{0.9}[1.0]{2006 September  19}\hspace*{-0.2cm}&\scalebox{0.9}[1.0]{OFF}\hspace*{-0.2cm}\\
\hspace*{-0.15cm}\scalebox{0.9}[1.0]{9}\hspace*{-0.2cm}&\scalebox{0.9}[1.0]{501071010}\hspace*{-0.2cm}&\scalebox{0.9}[1.0]{17 12 17.6, $-$39 18 50}\hspace*{-0.2cm}&\hspace*{-0.2cm}\scalebox{0.9}[1.0]{21}\hspace*{-0.2cm}&\hspace*{-0.2cm}\scalebox{0.9}[1.0]{2006 September  20}\hspace*{-0.2cm}&\scalebox{0.9}[1.0]{OFF}\hspace*{-0.2cm}\\
\hspace*{-0.15cm}\scalebox{0.9}[1.0]{10}\hspace*{-0.2cm}&\scalebox{0.9}[1.0]{501072010}\hspace*{-0.2cm}&\scalebox{0.9}[1.0]{17 15 44.5, $-$39 40 10}\hspace*{-0.2cm}&\hspace*{-0.2cm}\scalebox{0.9}[1.0]{20}\hspace*{-0.2cm}&\hspace*{-0.2cm}\scalebox{0.9}[1.0]{2006 October   \phantom{0}\phantom{0}\phantom{0}5}\hspace*{-0.2cm}&\scalebox{0.9}[1.0]{OFF}\hspace*{-0.2cm}\\
\hspace*{-0.15cm}\scalebox{0.9}[1.0]{11}\hspace*{-0.2cm}&\scalebox{0.9}[1.0]{504027010}\hspace*{-0.2cm}&\scalebox{0.9}[1.0]{17 11 50.8, $-$39 31 00}\hspace*{-0.2cm}&\hspace*{-0.2cm}\scalebox{0.9}[1.0]{62}\hspace*{-0.2cm}&\hspace*{-0.2cm}\scalebox{0.9}[1.0]{2010 February\phantom{0}\phantom{0}15}\hspace*{-0.2cm}&\scalebox{0.9}[1.0]{ON}\hspace*{-0.2cm}\\
\hspace*{-0.15cm}\scalebox{0.9}[1.0]{12}\hspace*{-0.2cm}&\scalebox{0.9}[1.0]{504028010}\hspace*{-0.2cm}&\scalebox{0.9}[1.0]{17 13 14.0, $-$40 14 22}\hspace*{-0.2cm}&\hspace*{-0.2cm}\scalebox{0.9}[1.0]{19}\hspace*{-0.2cm}&\hspace*{-0.2cm}\scalebox{0.9}[1.0]{2010 February\phantom{0}\phantom{0}16}\hspace*{-0.2cm}&\scalebox{0.9}[1.0]{ON}\hspace*{-0.2cm}\\
\hspace*{-0.15cm}\scalebox{0.9}[1.0]{13}\hspace*{-0.2cm}&\scalebox{0.9}[1.0]{504029010}\hspace*{-0.2cm}&\scalebox{0.9}[1.0]{17 12 39.8, $-$40 01 50}\hspace*{-0.2cm}&\hspace*{-0.2cm}\scalebox{0.9}[1.0]{21}\hspace*{-0.2cm}&\hspace*{-0.2cm}\scalebox{0.9}[1.0]{2010 February\phantom{0}\phantom{0}17}\hspace*{-0.2cm}&\scalebox{0.9}[1.0]{ON}\hspace*{-0.2cm}\\
\hspace*{-0.15cm}\scalebox{0.9}[1.0]{14}\hspace*{-0.2cm}&\scalebox{0.9}[1.0]{504030010}\hspace*{-0.2cm}&\scalebox{0.9}[1.0]{17 15 39.0, $-$40 00 47}\hspace*{-0.2cm}&\hspace*{-0.2cm}\scalebox{0.9}[1.0]{22}\hspace*{-0.2cm}&\hspace*{-0.2cm}\scalebox{0.9}[1.0]{2010 February\phantom{0}\phantom{0}17}\hspace*{-0.2cm}&\scalebox{0.9}[1.0]{ON}\hspace*{-0.2cm}\\
\hspace*{-0.15cm}\scalebox{0.9}[1.0]{OFF1}\hspace*{-0.2cm}&\scalebox{0.9}[1.0]{100026020}\hspace*{-0.2cm}&\scalebox{0.9}[1.0]{17 09 31.9, $-$38 49 24}\hspace*{-0.2cm}&\hspace*{-0.2cm}\scalebox{0.9}[1.0]{35}\hspace*{-0.2cm}&\hspace*{-0.2cm}\scalebox{0.9}[1.0]{2005 September 25}\hspace*{-0.2cm}&\scalebox{0.9}[1.0]{OFF}\hspace*{-0.2cm}\\
\hspace*{-0.15cm}\scalebox{0.9}[1.0]{OFF2}\hspace*{-0.2cm}&\scalebox{0.9}[1.0]{100026030}\hspace*{-0.2cm}&\hspace*{-0.2cm}\scalebox{0.9}[1.0]{17 09 05.1, $-$41 02 07}\hspace*{-0.2cm}&\hspace*{-0.2cm}\scalebox{0.9}[1.0]{38}\hspace*{-0.2cm}&\hspace*{-0.2cm}\scalebox{0.9}[1.0]{2005 September 28}\hspace*{-0.2cm}&\scalebox{0.9}[1.0]{OFF}\hspace*{-0.2cm}
\enddata
\label{tab1}
\tablecomments{The details of ID from 0 to 10, OFF1 and OFF2 are also shown in \cite{takahashi2008} and \cite{tanaka2008}.}
\end{deluxetable}

Aiming at better understanding results of X- and $\gamma$-ray observations and their relationship with the ISM, we are promoting a comprehensive comparison among the $\gamma$-rays, the X-rays, and the ISM in $\gamma$-ray SNRs (e.g., \citealp{fukui2008}; Paper I; \citealp {fukui2012}, hereafter paper II; \citealp{yoshiike2013}; \citealp{fukui2013}). Paper II presented an analysis of the interstellar protons both in molecular and atomic forms and showed a good spatial correspondence between the clumpy ISM protons and the VHE $\gamma$-rays. This provides crucial support for that the hadronic origin of the $\gamma$-rays is dominant instead of the leptonic origin. An alternative idea favoring the leptonic origin of the $\gamma$-rays was discussed recently in the context of the hard GeV--TeV $\gamma$-ray spectrum which appears similar to that expected in the leptonic origin  \citep[\citealp{abdo2011}; see also][]{ellison2010, ellison2012}. It is however argued that the hard spectrum is explained also in the hadornic origin. This is because only higher energy CR protons can interact with the dense gas to produce $\gamma$-rays due to energy-dependent penetration of CR protons into the dense gas \citep{zirakashvili2010}, which produces a spectrum as hard as that observed in RX J1713.7$-$3946 \citep{inoue2012}. It is also to be noted that the shock-cloud interaction does not favor the leptonic origin, because the interaction amplifies magnetic field to around 100 $\mu$G or even higher, which suppresses significantly the CR electrons via rapid synchrotron cooling and, accordingly, the leptonic component of $\gamma$-rays \citep[e.g.,][]{tanaka2008}.

\begin{figure*}
\begin{center}
\includegraphics[width=170mm,clip]{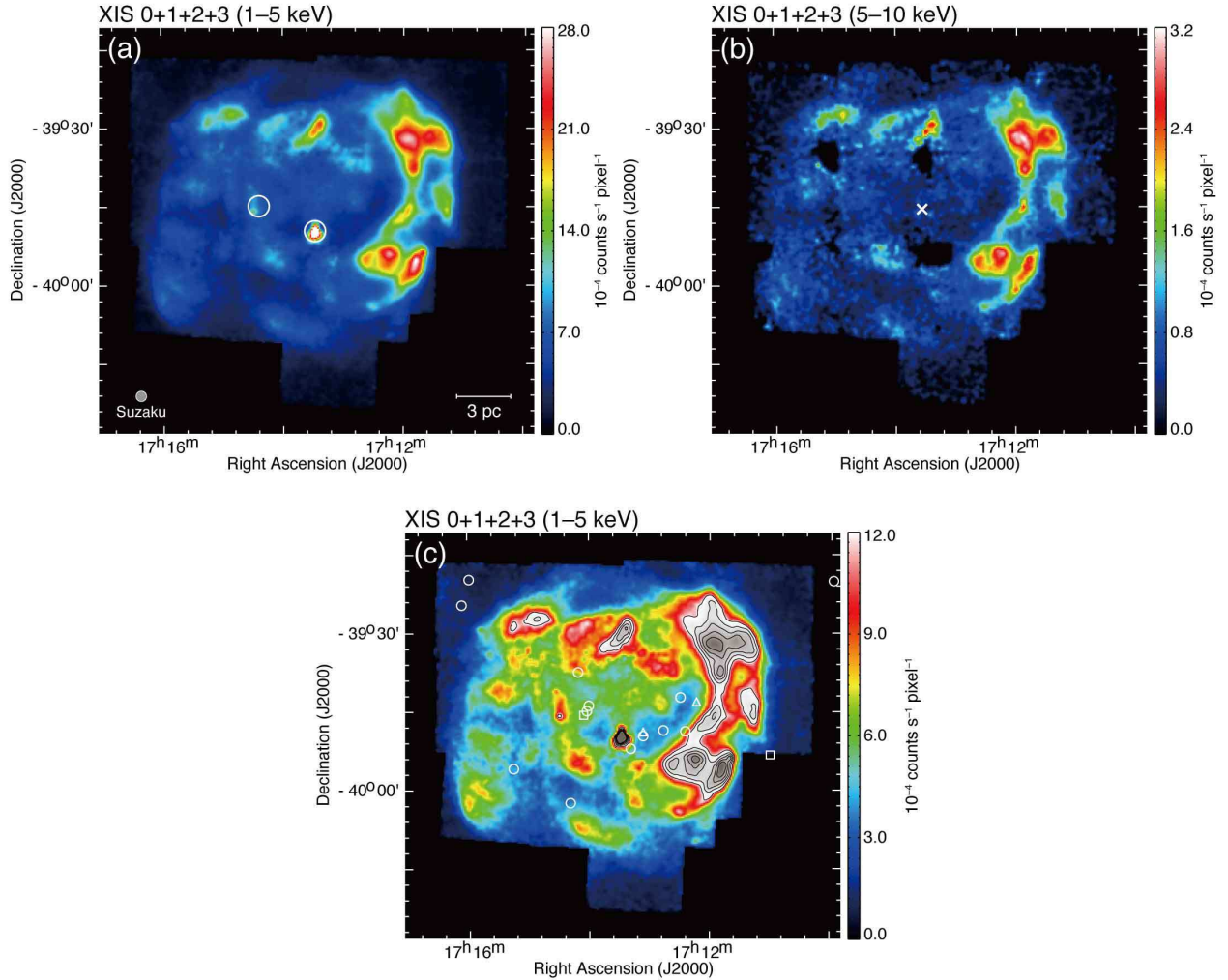}
\caption{$Suzaku$ XIS (XIS 0+1+2+3) mosaic images of RX J1713.7$-$3946 in the energy bands (a) 1--5 keV and (b) 5--10 keV. The color scale indicates the count rate on a linear scale. The color bar numbers are in units of 10$^{-4}$ counts s$^{-1}$ pixel$^{-1}$ with a pixel size of $\sim$16.7$\arcsec$. Both images are smoothed with a Gaussian kernel with FWHM of 45$''$. The positions of the two point-like sources are shown with large circles in (a) (see Table \ref{tab2}). (c) Same XIS mosaic image (1--5 keV) as (a), but the color scale is changed to emphasize the region of low photon counts below 12 $\times$ 10$^{-4}$ counts s$^{-1}$ pixel$^{-1}$. Above this level shown in gray scale, the lowest contour level and the contour interval are 12 and 4 $\times$ 10$^{-4}$ counts s$^{-1}$ pixel$^{-1}$, respectively. The small circles, triangles and squares show the position of seven X-rays point source, three pulsars and two Wolf-Rayet stars, respectively.}
\label{fig2}
\end{center}
\end{figure*}%

\begin{deluxetable*}{llllllc}
\tabletypesize{\scriptsize}
\tablecaption{Summary of the X-ray point sources toward RX J1713.7$-$3946}
\vspace*{-0.8cm}
\tablehead{&&\multicolumn{1}{c}{$\alpha_{\mathrm{J2000}}$} & \multicolumn{1}{c}{$\delta_{\mathrm{J2000}}$} &&&\\
\multicolumn{1}{c}{Name}& &\multicolumn{1}{c}{($^{\mathrm{h}}$ $^{\mathrm{m}}$ $^{\mathrm{s}}$)} & \multicolumn{1}{c}{($^{\circ}$ $\arcmin$ $\arcsec$)}&&\multicolumn{1}{c}{Source Type}&References}
\startdata
WR 84 &&17 11 21.70\phantom{0}& $-$39 53 22.2\phantom{0}&& Wolf-Rayet star$\dag$&1\\
CD$-$39 11212B&&17 14 27.129& $-$39 45 47.25&&Wolf-Rayet star&2\\
PSR J1712$-$3943A&&17 12 35.0\phantom{0}\phantom{0}&$-$39 43 14\phantom{0}\phantom{0}&&pulsar&3\\
PSR J1712$-$3943B&&17 12 35.0\phantom{0}\phantom{0}&$-$39 43 14\phantom{0}\phantom{0}&&pulsar&3\\
PSR J1713$-$3949&&17 13 28\phantom{0}\phantom{0}\phantom{0}&$-$39 49.0\phantom{0}\phantom{0}\phantom{0}\phantom{0}&&pulsar$\dag$&4\\
EXMS B1709$-$397A&&17 12 46\phantom{0}\phantom{0}\phantom{0}&$-$39 48.9\phantom{0}\phantom{0}\phantom{0}\phantom{0}&&X-ray point source&5\\
GPS 1709$-$396&&17 12 51.0\phantom{0}\phantom{0}&$-$39 42 25\phantom{0}\phantom{0}&&X-ray point source&6\\
EXMS B1709$-$397B&&17 13 08\phantom{0}\phantom{0}\phantom{0}&$-$39 48.7\phantom{0}\phantom{0}\phantom{0}\phantom{0}&& X-ray point source&5\\
1WGA J1713.4$-$3949&&17 13 28\phantom{0}\phantom{0}\phantom{0}&$-$39 49.8\phantom{0}\phantom{0}\phantom{0}\phantom{0}&&X-ray point source$\dag$&4, 7\\
CXOPS J171340.5$-$395213&&17 13 40.5\phantom{0}\phantom{0}&$-$39 52 13\phantom{0}\phantom{0}&&X-ray point source&8\\
EXO 1710$-$396&&17 14 22\phantom{0}\phantom{0}\phantom{0}&$-$39 44.0\phantom{0}\phantom{0}\phantom{0}\phantom{0}&&X-ray point source$\ddag$&9\\
1WGA J1714.4$-$3945&&17 14.4\phantom{0}\phantom{0}\phantom{0}\phantom{0}\phantom{0}&$-$39 45\phantom{0}\phantom{0}\phantom{0}\phantom{0}\phantom{0}&&X-ray point source$\ddag$&10, 11
\enddata
\label{tab2}
\tablecomments{$\dag$ and $\ddag$ sources are connected with two X-ray point-like sources shown in Figure \ref{fig2}, respectively.}
\tablerefs{(1) \citealp{van2001}; (2) \citealp{cutri2003}; (3) \citealp{burgay2006}; (4) \citep{lazendic2003}; (5) \citealp{reynolds1999}; (6) \citealp{gottwald1995}; (7) \citealp{landt2008}; (8) \citealp{van2012}; (9) \citealp{lu1999}; (10) \citealp{slane1999}; (11) \citealp{pfeffermann1996}.}
\end{deluxetable*}

\begin{figure*}
\begin{center}
\tabletypesize{\scriptsize}
\includegraphics[width=180mm,clip]{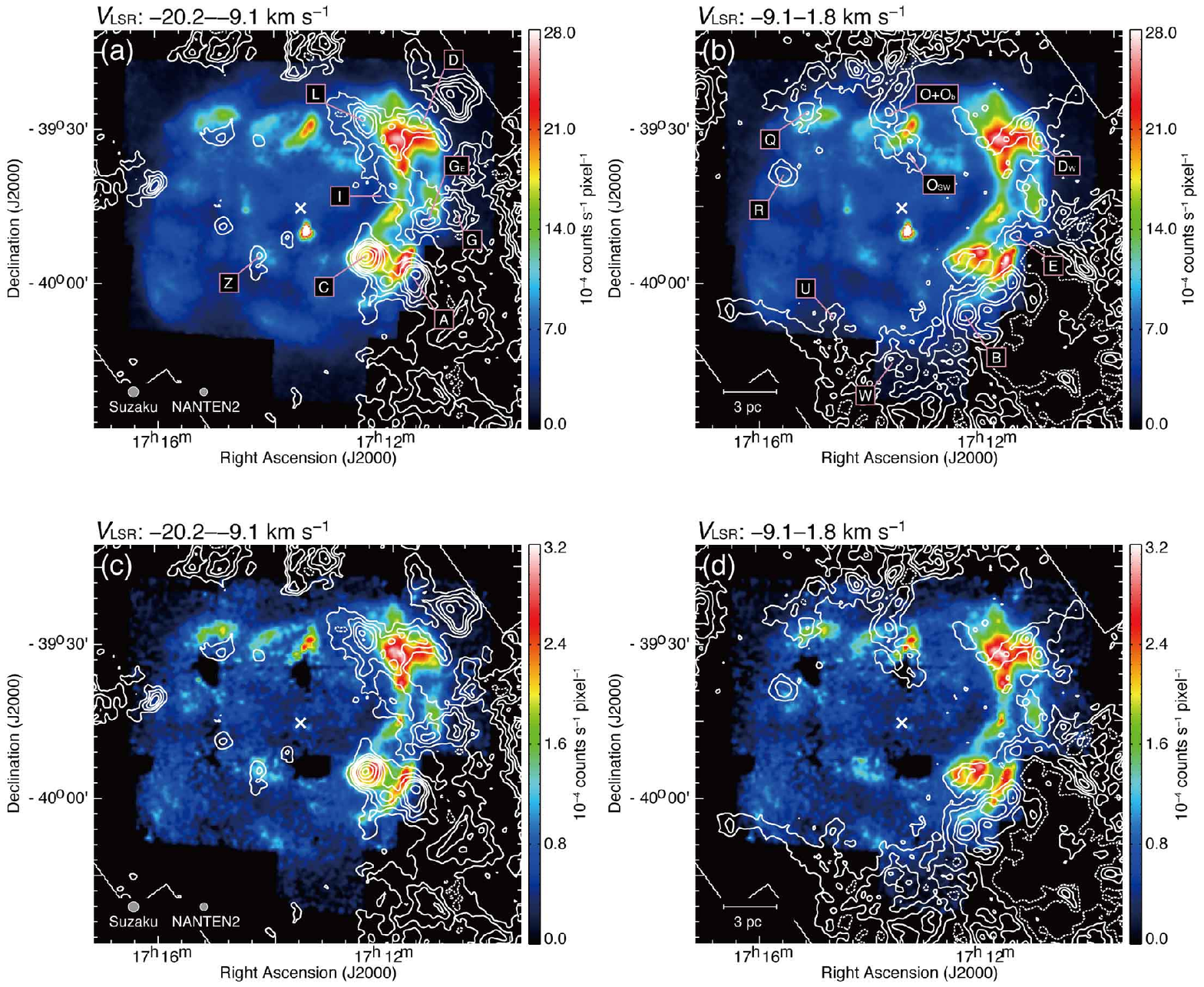}
\caption{Pair of $^{12}$CO($J$=2--1) velocity channel maps ($white$ $contours$) superposed on the $Suzaku$ XIS mosaic image in two energy bands (a, b: 1--5 keV and, c, d: 5--10 keV) in color scale. The velocity ranges are (a, c) $-20.2$--$-9.1$ km s$^{-1}$ and (b, d) $-9.1$--$1.8$ km s$^{-1}$, respectively. The lowest contour level and the contour interval of CO are 3.1 K km s$^{-1}$ ($\sim3\sigma$) in (b) and (d). In (a) and (c), the contour level are 3.1, 6.2, 9.3, 12.4, 15.5, 21.7, 27.9, 31.0 K km s$^{-1}$. The CO clumps discussed in Section \ref{section:analysis} are indicated in the figure.}
\label{fig3}
\end{center}
\end{figure*}%

Now, in RX J1713.7$-$3946, the next step is to establish the connection between the synchrotron X-rays and the inhomogeneous ISM distribution {\it{over the entire SNR}}. The present work is aimed at better understanding the SNR shock-cloud interaction and establishing thereby the origin of the distribution of the synchrotron X-rays in the SNR. This study will be extended to the other SNRs with non-thermal features, allowing us to deepen our understanding on the role of the interacting ISM in the high energy radiation and in the CR acceleration. In the present paper we show a comparison of the spatial distribution among CO, H{\sc i} and X-rays over the whole SNR in order to clarify the relationship between dense gas and the high-energy electrons. We will publish a separate paper that deals with a detailed spectral analysis of the $Suzaku$ X-ray results compared with the interstellar gas distribution (Sano et al. 2013b in prep.).

The paper is organized as follows. Section \ref{section:datasets} gives the description of the datasets of CO, H{\sc i} and X-rays and Section \ref{section:analysis} gives the analysis. Section \ref{section:discussion} gives discussion and Section \ref{section:conclusions} the conclusions.

\begin{deluxetable*}{lcccccccccccc}
\tabletypesize{\scriptsize}
\tablecaption{Properties of CO Clumps}
\tablehead{&\multicolumn{6}{c}{$^{12}$CO($J$=1--0)}&&\multicolumn{5}{c}{$^{12}$CO($J$=2--1)}\\
\cline{2-7}\cline{9-13}
\multicolumn{1}{c}{Name} & $\alpha_{\mathrm{J2000}}$ & $\delta_{\mathrm{J2000}}$ & $T_{\rm R^\ast} $ & $V_{\mathrm{peak}}$ & $\Delta V_{\mathrm{LSR}}$ & Mass&& $\alpha_{\mathrm{J2000}}$ & $\delta_{\mathrm{J2000}}$ & $T_{\rm R^\ast} $ & $V_{\mathrm{peak}}$ & $\Delta V_{\mathrm{LSR}}$\\
& ($^{\mathrm{h}}$ $^{\mathrm{m}}$ $^{\mathrm{s}}$) & ($^{\circ}$ $\arcmin$ $\arcsec$) & (K) & \scalebox{0.9}[1]{(km $\mathrm{s^{-1}}$)} & \scalebox{0.9}[1]{(km $\mathrm{s^{-1}}$)} & ($M_\sun $) & & ($^{\mathrm{h}}$ $^{\mathrm{m}}$ $^{\mathrm{s}}$) & ($^{\circ}$ $\arcmin$ $\arcsec$) & (K) & \scalebox{0.9}[1]{(km $\mathrm{s^{-1}}$)} & \scalebox{0.9}[1]{(km $\mathrm{s^{-1}}$)} \\
\multicolumn{1}{c}{(1)} & (2) & (3) & (4) & (5) & (6) & (7) & &(8) & (9) & (10) & (11) & (12)}
\startdata
A........ &  17 11 35.9 & $-39$ 59 01.8 & 8.5 & $-10.3$ & 4.8& 686 && 17 11 38.4 & $-39$ 58 46.9 & 6.6 & $-10.0$ & 4.5 \\
B........ &  17 12 26.5 & $-40$ 06 06.3 & 4.2 & \phantom{0}$-8.0$ & 4.6 & 190 && 17 12 26.8 & $-40$ 05 55.5 & 3.3 & \phantom{0}$-8.1$ & 4.5 \\
C........ &  17 12 25.9 & $-39$ 56 04.4 & 9.4 & $-12.0$ & 3.8 & 397 && 17 12 27.0 & $-39$ 54 58.0 & 7.5 & $-11.9$ & 4.6  \\
C\tiny{\textsc{E}}....... & 17 13 01.3 & $-39$ 53 35.2 & 1.1 & \phantom{0}$-9.1$ & 1.6 & \phantom{0}10 & & 17 12 57.4 & $-$39 53 42.3 & 3.7 & \phantom{0}$-$8.8& 1.4 \\
D........ &  17 11 28.0 & $-39$ 30 37.6  & 4.0 & $-11.1$ & 4.8 & 292&  & 17 11 32.5 & $-39$ 30 03.9 & 3.3 & \phantom{0}$-9.3$ & 4.8 \\
D\tiny{\textsc{W}}...... & 17 11 01.3 & $-39$ 34 17.6 & 2.3 & \phantom{0}\phantom{0}2.4 & 3.3 & 137 & & 17 11 34.3 & $-39$ 32 31.2 & 3.1 & \phantom{0}$-1.1$ & 6.0  \\
E........ &  17 11 29.1 & $-39$ 50 38.5 & 2.0 & \phantom{0}$-6.1$ & 7.2 & 159 && 17 11 55.4 & $-39$ 51 07.2  & 2.0 & \phantom{0}$-6.0$ & 5.0  \\
G........&  17 10 55.6 & $-39$ 45 55.2 & 3.3 & $-10.8$ & 8.0 &  307&& 17 10 56.3 & $-39$ 45 20.6 & 2.8 & $-11.5$ & 4.8  \\
G\tiny{\textsc{E}}....... &  17 11 27.1 & $-39$ 47 49.6  & 5.4 & $-12.8$ & 2.6 & 168 && 17 11 21.0 & $-39$ 47 24.4 & 4.3 & $-12.3$ & 2.7  \\
I......... &  17 12 08.2 & $-39$ 43 43.3 & 1.8 & \phantom{0}$-9.9$ & 5.4 & 103 & & 17 12 16.6 & $-39$ 43 22.8 & 1.3 & $-10.4$ & 5.9 \\
L........ &  17 12 25.8 & $-39$ 28 53.4 & 4.0 & $-12.0$ & 5.7 &  370 && 17 12 30.2 & $-39$ 28 14.6 & 3.2 & $-11.7$ & 6.0 \\
O........ &  17 13 46.7 & $-39$ 27 49.8  & 1.1 & \phantom{0}$-6.4$ & 4.9 & \phantom{0}61 && 17 13 46.0 & $-39$ 26 28.6 & 1.2 & \phantom{0}$-4.6$ & 4.7\\
O{\tiny{b}}...... &  17 13 46.7 & $-39$ 27 49.8  & 1.9 &  \phantom{0}\phantom{0} 1.1 & 3.4 & \phantom{0}80  & & 17 13 46.0 & $-39$ 26 28.6 & 1.9 & \phantom{0}\phantom{0} 1.0 & 3.8  \\
O\tiny{\textsc{SW}}.... & 17 13 24.8 & $-39$ 37 08.6 & 2.8 & \phantom{0}$-1.6$ & 2.0 & \phantom{0}60 && 17 13 08.9 & $-39$ 36 43.2 & 3.5 & \phantom{0}$-1.3$ & 1.3 \\
Q........ & 17 15 13.4 & $-39$ 25 06.2 & 2.9 & \phantom{0}$-2.8$ & 3.2 & 108& & 17 15 11.7 & $-39$ 26 47.4 & 2.8 & \phantom{0}$-2.2$ & 2.8\\
Q\tiny{\textsc{W}}...... & 17 14 49.3 & $-39$ 31 35.1 & 3.0 & $-14.3$ & 2.4 & \phantom{0}46 & & 17 14 53.0 & $-$39 31 30.7 & 3.7 & $-$14.1 & 1.8  \\
R........  & 17 15 39.9 & $-39$ 38 34.6 & 4.1 & \phantom{0}$-3.3$ & 2.4 &\phantom{0}67 && 17 15 32.0 & $-39$ 39 28.5 & 3.2 & \phantom{0}$-3.1$ & 2.2 \\
U........ & 17 14 34.2 & $-40$ 06 27.0 & 3.7 & \phantom{0}$-4.8$ & 1.3 & \phantom{0}58 & &  17 14 13.4 &  $-40$ 06 25.2 & 3.0 & \phantom{0}$-4.6$ & 1.3 \\
W....... & 17 13 42.8 & $-40$ 16 40.7 & 5.0 & \phantom{0}$-5.1$ & 3.0 & 402 & &  17 13 30.0 & $-40$ 15 03.6 & 3.3 & \phantom{0}$-4.9$ & 3.4 \\
Z........ & 17 14 18.7 & $-39$ 56 55.1  & 2.6 & $-20.1$ & 2.7 & \phantom{0}72 &&  17 13 53.3 & $-39$ 54 46.8  & 2.9 & $-19.8$ & 2.9  \\
Z\tiny{\textsc{NW}}.... & 17 13 45.0 & $-39$ 52 14.9  & 3.0 &$-19.8$ & 2.6 & \phantom{0}36  && 17 13 45.6 & $-$39 51 09.2 & 5.1 & $-$19.8 & 1.8  \\
Z\tiny{\textsc{NE}}..... & 17 14 57.4 & $-39$ 49 59.3 & 2.2 & $-20.0$ & 3.3 & \phantom{0}31 & & 17 14 53.3 & $-$39 49 25.7 & 3.6 & $-$19.6 & 2.5
\enddata
\label{tab3}
\tablecomments{\tiny{Col. (1): Clump name. Cols. (2--7) and (8--12): Observed properties of the $^{12}$CO($J$=1--0, 2--1) spectra obtained at the peak positions of the CO clumps. Cols. (2)--(3): Position of the peak CO intensity. Col. (4): Peak radiation temperature $T_{\rm R^{\ast}} $. Col. (5): $V_{\mathrm{peak}}$ derived from a single Gaussian fitting. Col. (6): FWHM line width $\bigtriangleup V_{\mathrm{LSR}}$. Col. (7): Total mass of the clumps derived by using the relation between the molecular hydrogen column density $N$($\mathrm{H_2}$) and the $^{12}$CO($J$=1--0) intensity $W$($^{12}$CO), $N$($\mathrm{H_2}$) = 2.0 $\times$ $10^{20}$[$W$($^{12}$CO) (K km $\mathrm{s^{-1}}$)] ($\mathrm{cm^{-2}}$) \citep{bertsch1993}. See also the text for more details. Cols. (8--12): The observed properties same with Cols. (2--6) for the $^{12}$CO($J$=2--1) spectra. The properties of A--C, D, E, G, I--O, Q and R--W derived from $^{12}$CO($J$=1--0) are shown by \cite{moriguchi2005, sano2010}.}}
\end{deluxetable*}

\begin{figure}
\begin{center}
\includegraphics[width=86mm,clip]{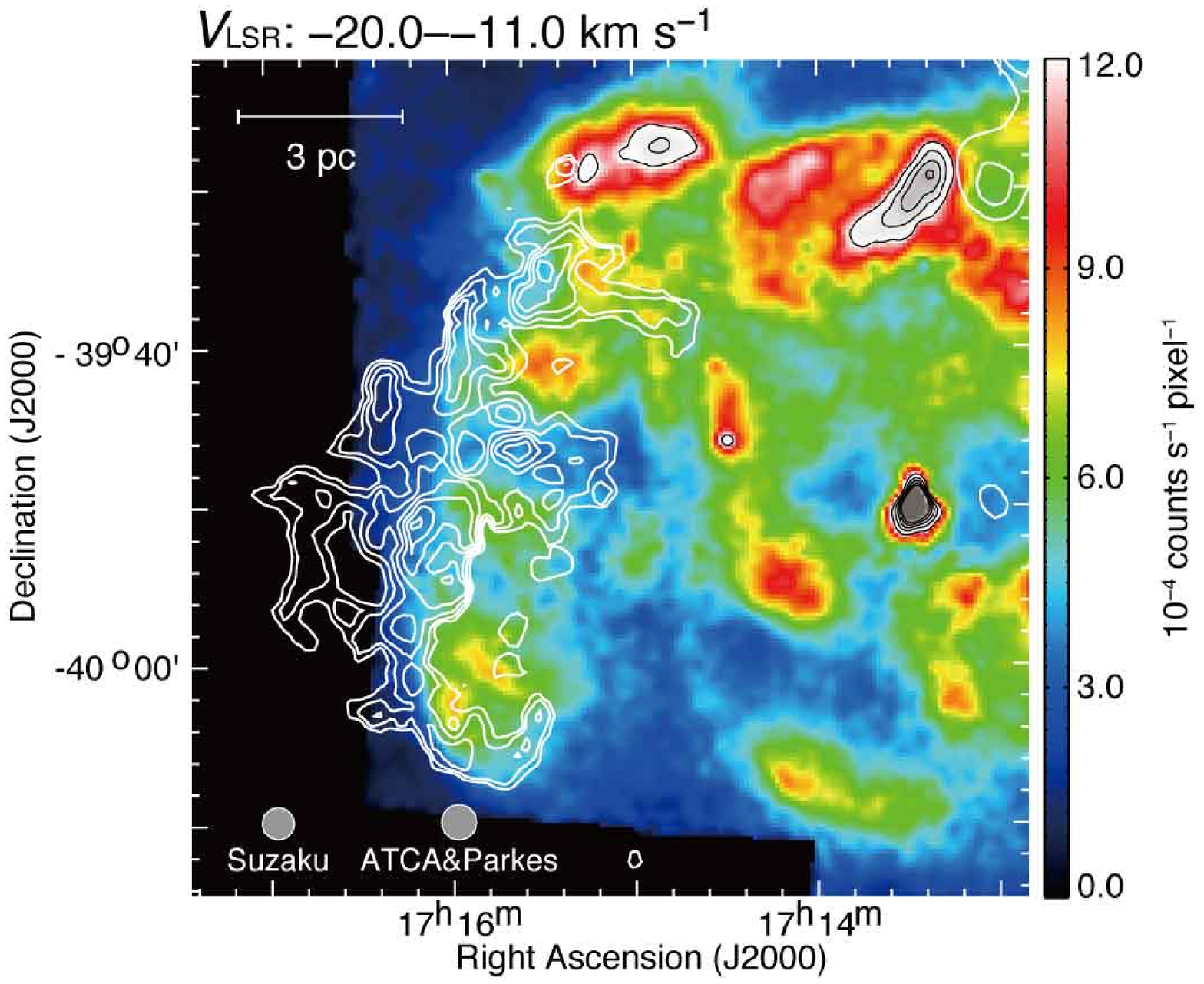}
\caption{Same XIS mosaic image (1--5 keV) as Figure \ref{fig2} (c) toward the the SE-rim. The white contours indicate the distribution of H{\sc i} proton column density (self-absorption corrected; see also the text and Paper II). The lowest contour level and the contour interval in H{\sc i} proton column density are 2.0 and 0.1 $\times$ 10$^{21}$ cm$^{-2}$, respectively. The velocity range is $-20.0$--$-11.0$ km s$^{-1}$.}
\label{fig4}
\end{center}
\end{figure}%

\vspace*{0.2cm}
\section{Observations and data reductions}\label{section:datasets}

\subsection{CO}
The $^{12}$CO($J$=2--1) data at 230 GHz were taken with the NANTEN2 4-m telescope in the period from August to November in 2008, and were published in Papers I, II and \citet{maxted2012}. The front end was a 4 K cooled double sideband (DSB) receiver and a typical system temperature was $\sim $250 K in the single sideband (SSB) including the atmosphere toward the zenith. The telescope had an angular resolution (FWHM) of $90\arcsec$ at 230 GHz. We used an acoustic optical spectrometer having 2048 channels with a bandwidth of 390 km s$^{-1}$ and resolution per channel of 0.38 km s$^{-1}$. Observations were carried out in the on-the-fly mode with an integration time of 1.0 s or 2.0 s per grid, and provided a Nyquist-sampled 30$\arcsec$ grid dataset. The ambient temperature load was employed for the intensity calibration. The absolute intensity scale was estimated by observing the Ori KL object [$5^{\mathrm{h}} 35^{\mathrm{m}} 14\fs52; -5{^\circ} 22\arcmin 28\farcs2$ (J2000)] \citep{schneider1998} and the main beam efficiency, $\eta_{\rm mb}$, was estimated to be 0.83. The rms noise fluctuations with 1.0 s and 2.0 s integrations were better than 0.66 K and 0.51 K per channel, respectively. The pointing accuracy was estimated to be better than $\sim$15$''$ by observing Jupiter every two hours. The image was smoothed with a Gaussian kernel with FWHM of 60$''$. The observed area is shown in Figure \ref{fig1}. In addition, we used the $^{12}$CO($J$=1--0) data at 115 GHz taken with the NANTEN 4-m telescope which were already published in \cite{moriguchi2005, fukui2012}. The angular resolution of the data was $2.6\arcmin$ (FWHM) and the velocity resolution and rms noise fluctuations are 0.65 km s$^{-1}$ and 0.3 K, respectively. Observations were carried out in the position-switching mode with a $2\arcmin$ grid spacing \citep[for more detailed information, see also][]{moriguchi2005,fukui2012}. The $^{12}$CO($J$=2--1) data were used for comparison with the X-ray images, while the $^{12}$CO($J$=1--0) data mainly for estimating molecular mass from the integrated $^{12}$CO intensity $W$($^{12}$CO) (K km s$^{-1}$) by using a relationship, $N$(H$_2$) = $X_{\mathrm{CO}}$ $\cdot$ [$W$($^{12}$CO) (K km s$^{-1}$)] (cm$^{-2}$), where an $X_{\mathrm{CO}}$ factor of 2.0 $\times$ 10$^{20}$ (cm$^{-2}$ (K km s$^{-1}$)$^{-1}$) is adopted \citep{bertsch1993}.

\begin{figure*}
\begin{center}
\includegraphics[width=182mm,clip]{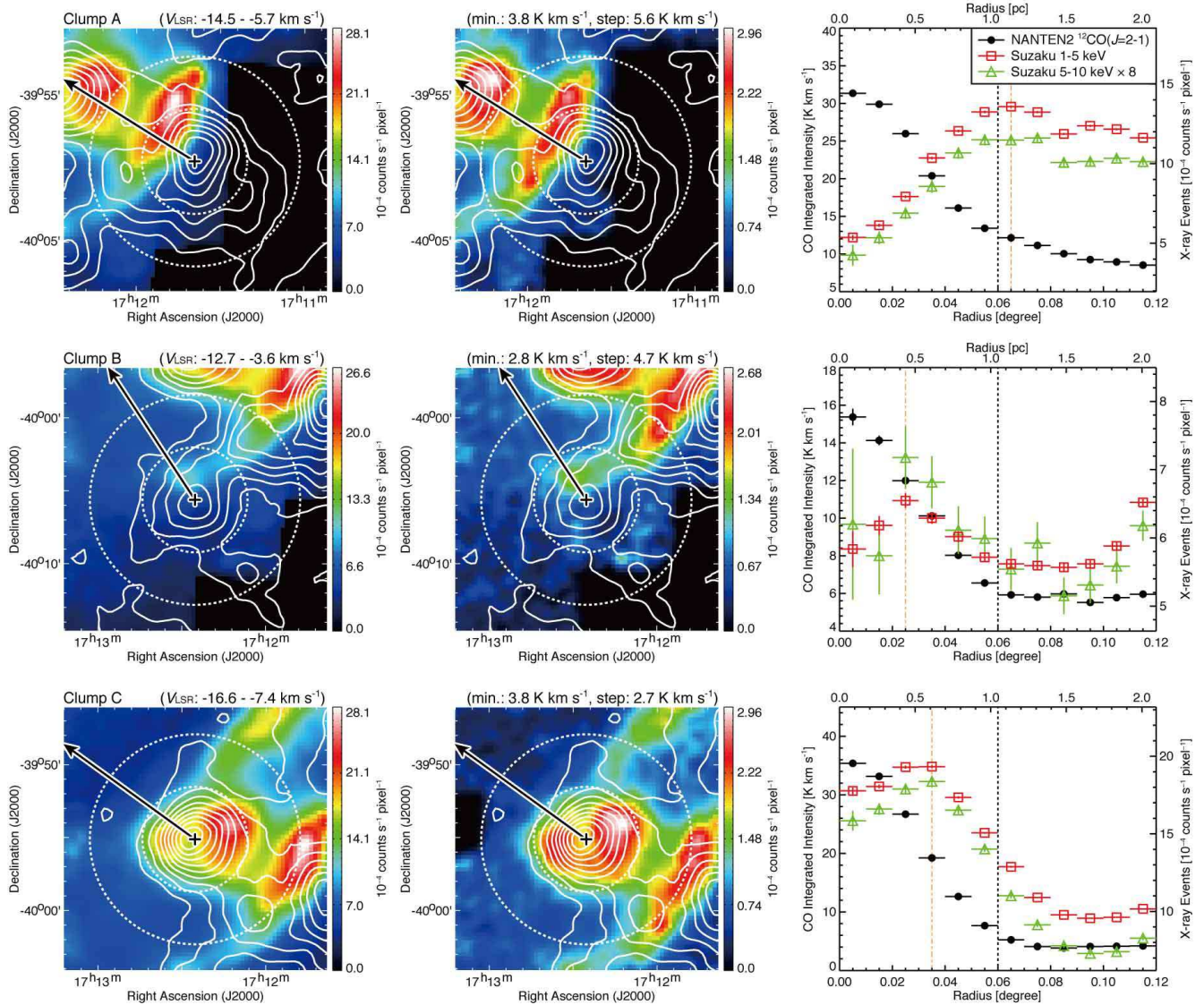}
\caption{\scriptsize{Distribution of $^{12}$CO($J$=2--1) emission ($white$ $contours$) superposed on the $Suzaku$  1--5 keV (left) and 5--10 keV (right) images. Velocity range in integration and contour levels are shown in the top of left and middle panels, respectively. Each arrow indicates the direction of the center of the SNR. The crosses show the position of the center of gravity for each CO clump (see also Table \ref{tab4}). The dashed white circles represent radii 0$\fdg$06 and 0$\fdg$12 of the center of gravity for each CO clump. Right panels show the radial profiles around each molecular clump in the $^{12}$CO($J$=2--1) integrated intensity and $Suzaku$ two energy bands (1--5 keV and 5--10 keV, in units of 10$^{-4}$ counts s$^{-1}$ pixel$^{-1}$) in Figure \ref{fig2}. The radial profiles from the 5--10 keV band has been scaled such that it has the same area as the 1--5 keV profile (scaled by a factor 8), for the sake of direct comparison. The orange dash-dotted lines indicate X-ray peak radius in the energy band 1--5 keV.}}
\label{fig5}
\end{center}
\end{figure*}

\begin{figure*}
\begin{center}
\figurenum{5}
\includegraphics[width=182mm,clip]{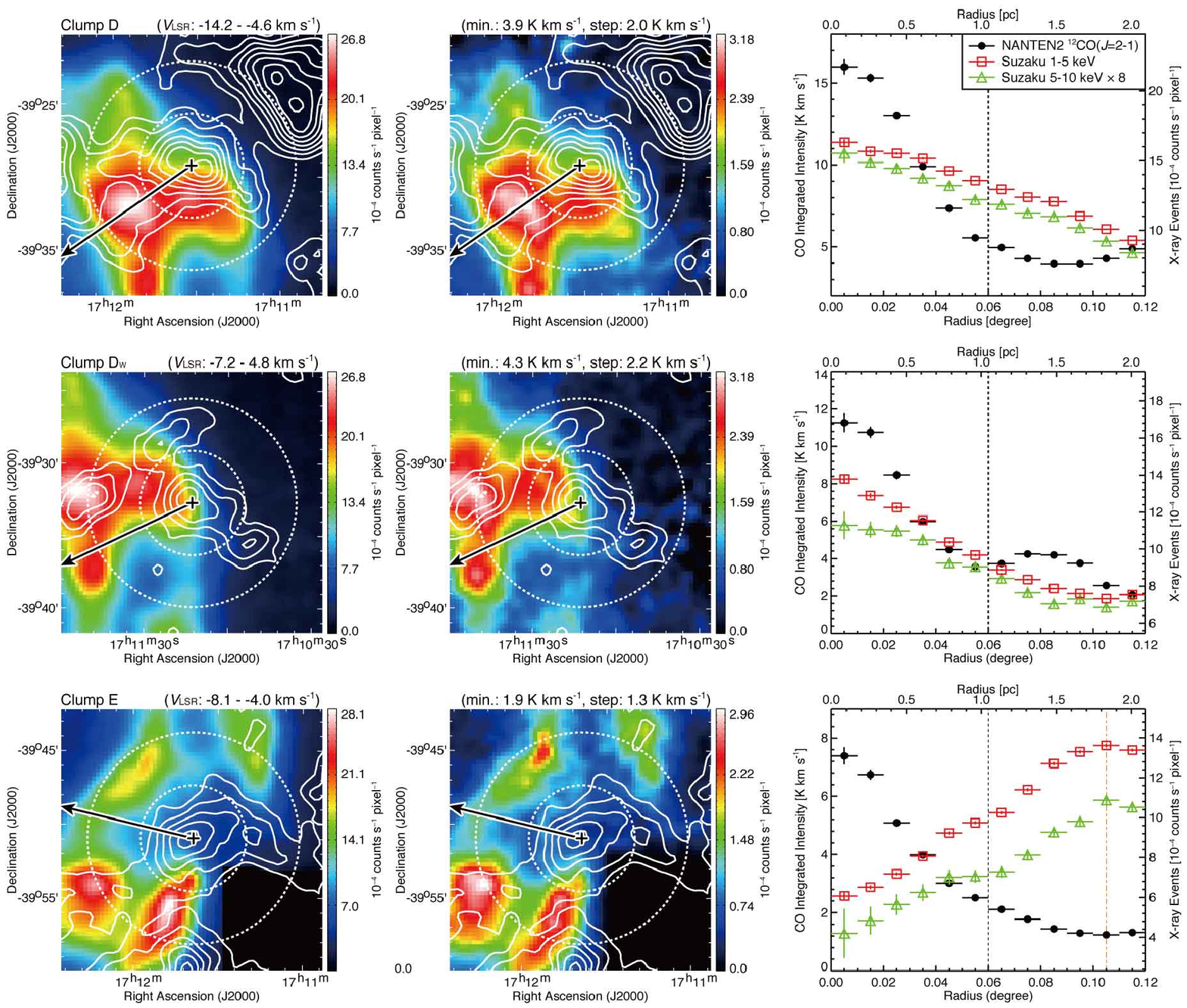}
\caption{$continued.$}
\end{center}
\end{figure*}%

\begin{figure*}
\begin{center}
\figurenum{5}
\includegraphics[width=182mm,clip]{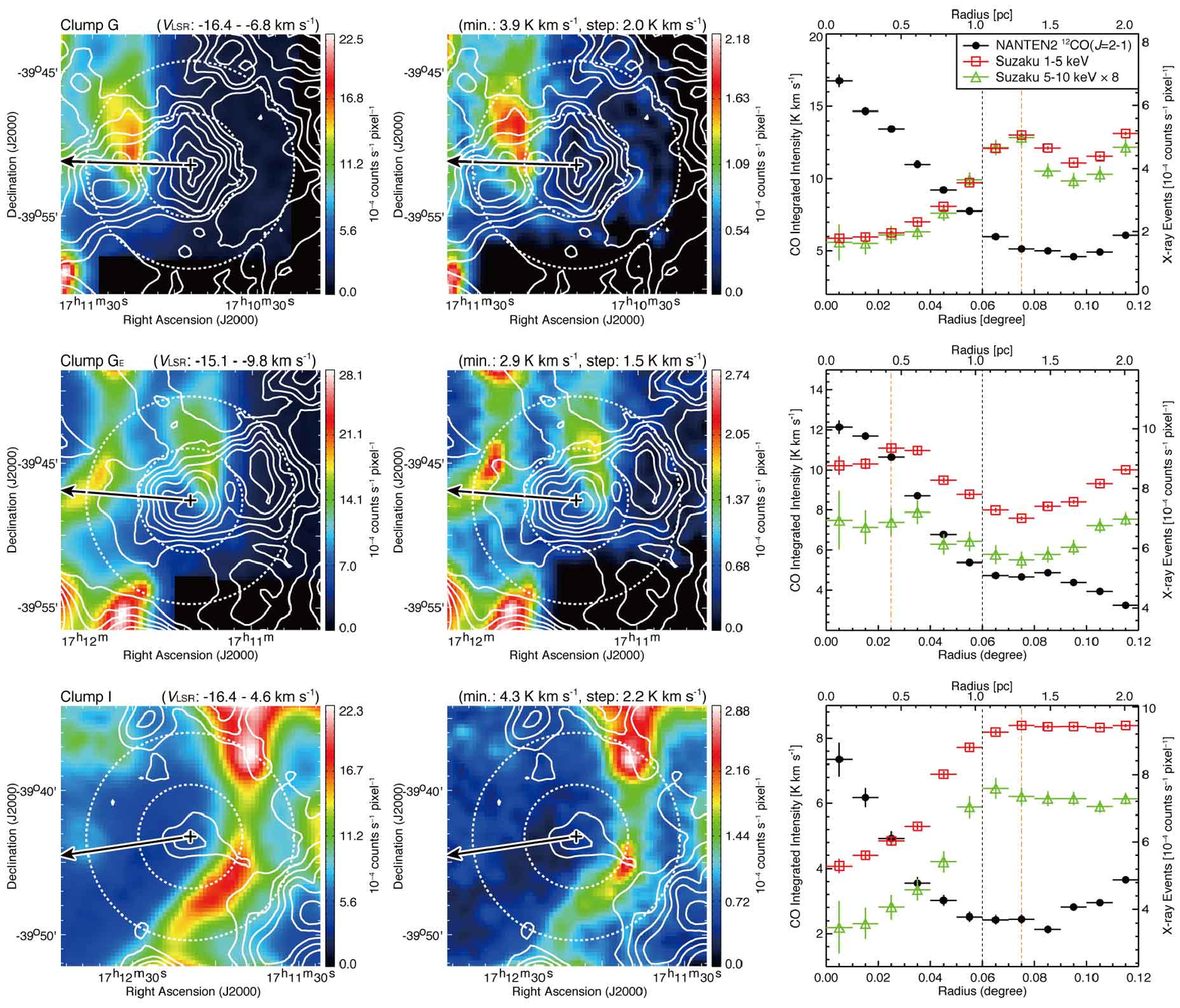}
\caption{$continued.$}
\end{center}
\end{figure*}%

\begin{figure*}
\begin{center}
\figurenum{5}
\includegraphics[width=182mm,clip]{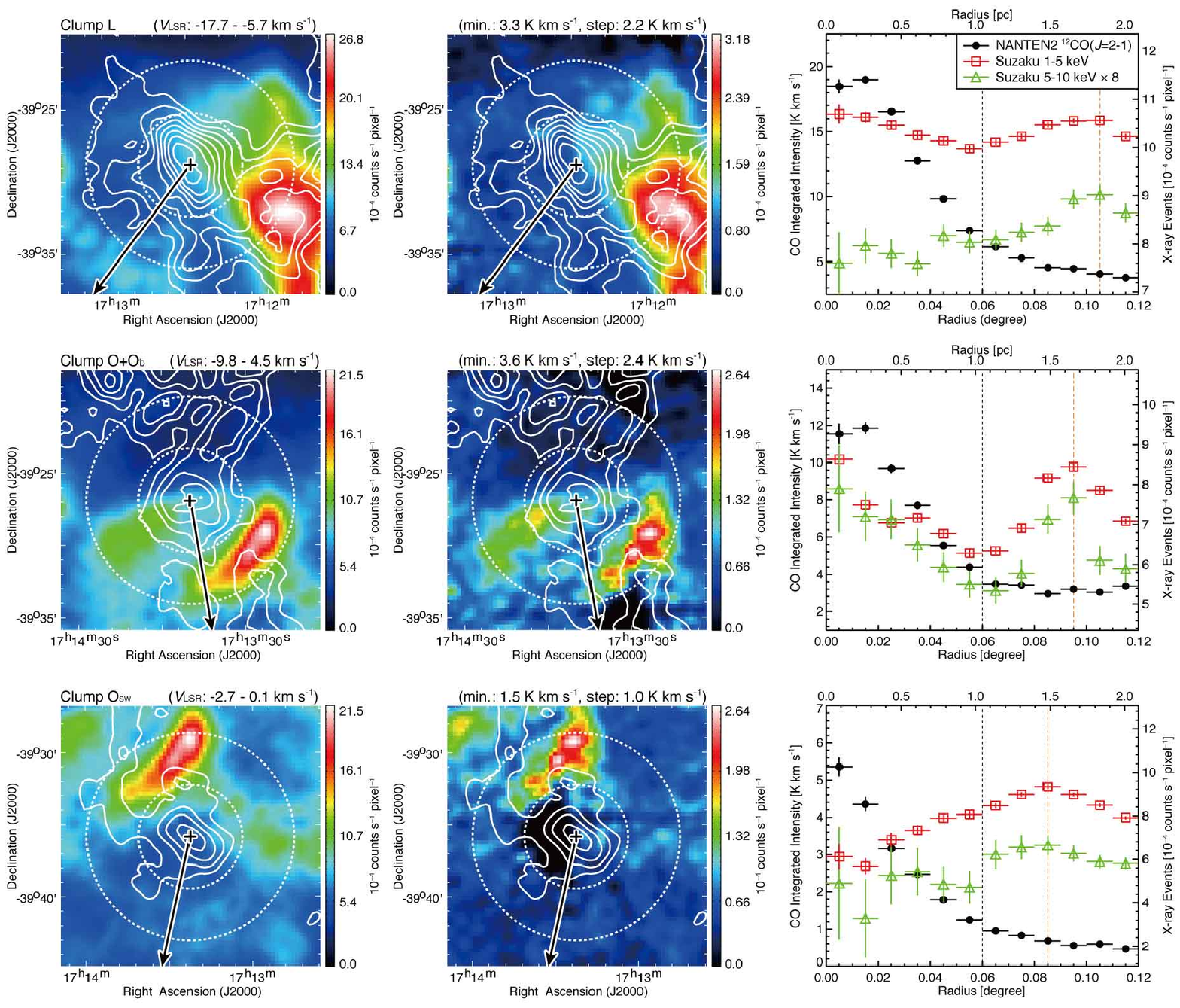}
\caption{$continued.$}
\end{center}
\end{figure*}%

\begin{figure*}
\begin{center}
\figurenum{5}
\includegraphics[width=182mm,clip]{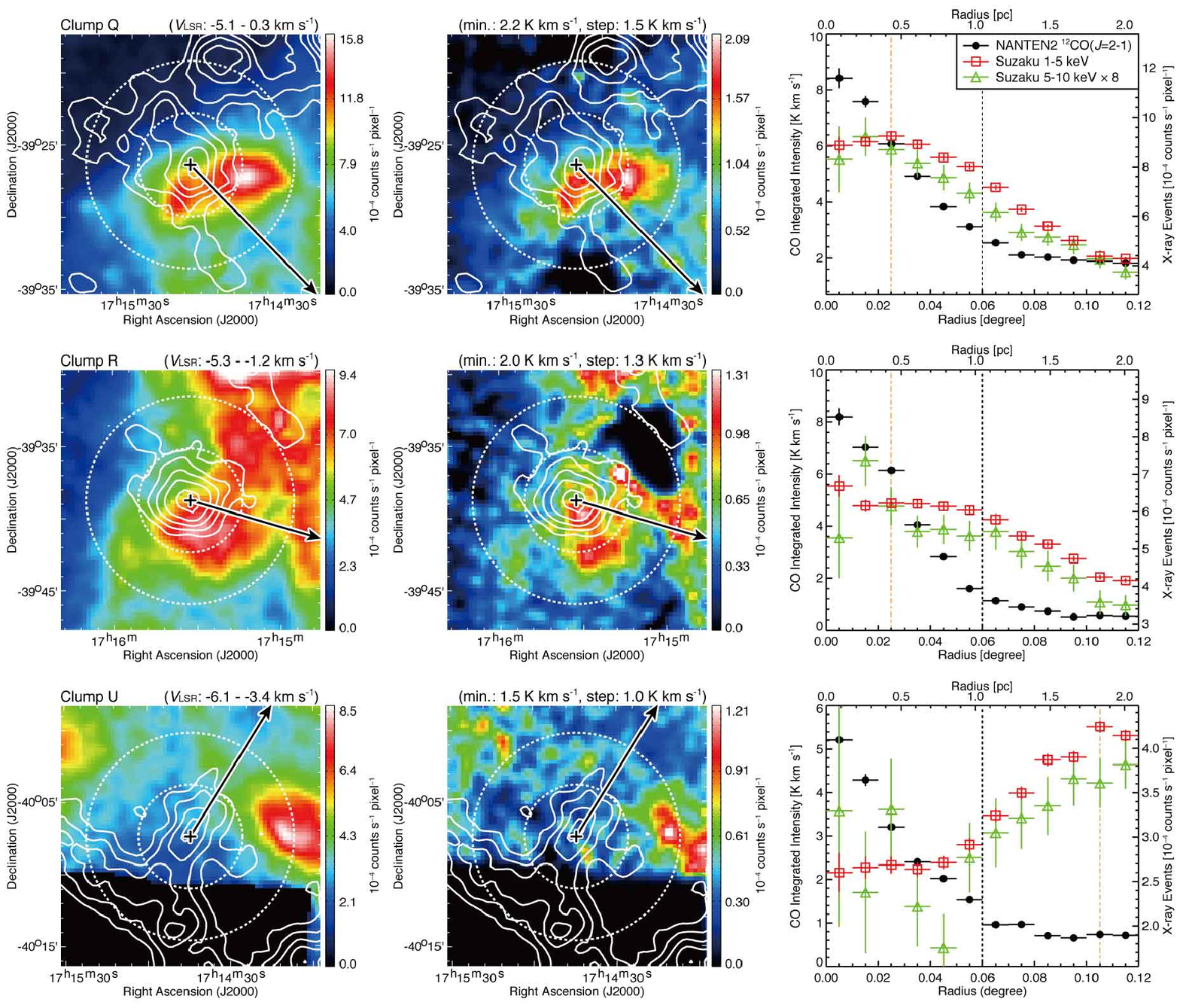}
\caption{$continued.$}
\end{center}
\end{figure*}%

\begin{figure*}
\begin{center}
\figurenum{5}
\includegraphics[width=182mm,clip]{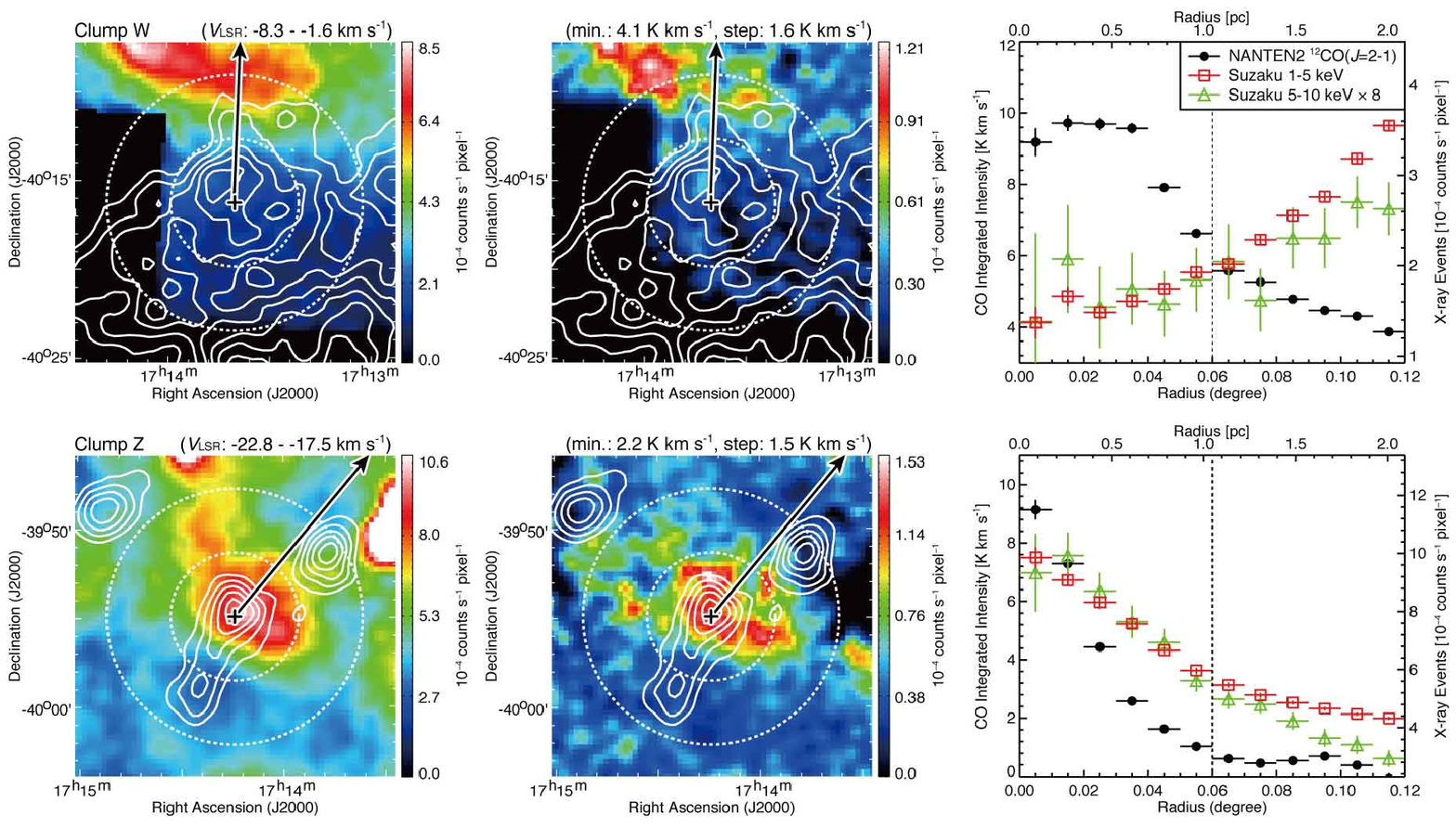}
\caption{$continued.$}
\end{center}
\end{figure*}%

\begin{figure}
\begin{center}
\includegraphics[width=80mm,clip]{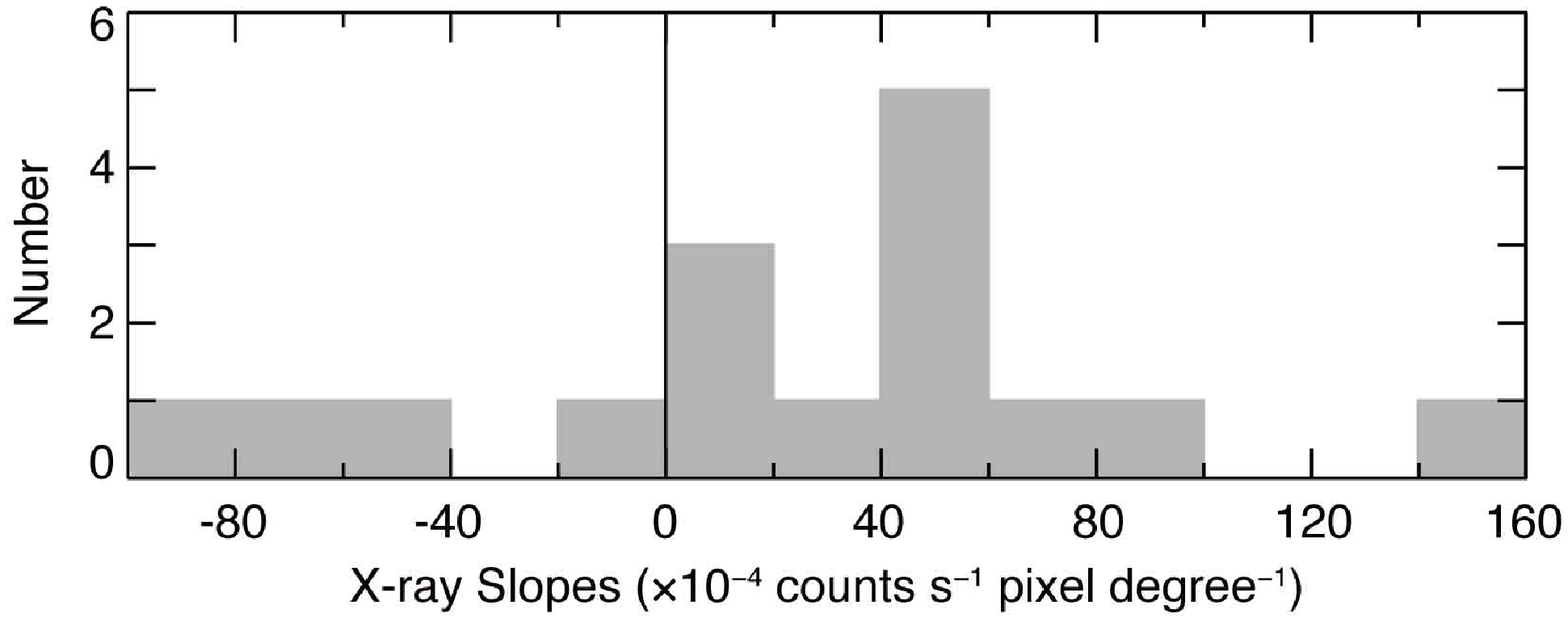}
\caption{Histogram of the X-ray slope estimated by a linear fitting in the radial profile for each CO clump. The fitting ranges and results are shown in Table \ref{tab5}.}
\label{fig6}
\end{center}
\end{figure}%

\subsection{H{\sc i}}
The 21 cm H{\sc i} spectral data were taken from the Southern Galactic Plane Survey \citep[SGPS;][]{mccluregriffiths2005} with the Australia Telescope Compact Array (ATCA) combined with the 64-m Parkes Radio Telescope. The combined beam size and the grid spacing of the dataset are 2.2$\arcmin$ and $40\arcsec$, respectively. The velocity resolution and typical rms noise fluctuations were 0.82 km s$^{-1}$ and 1.9 K, respectively. We applied the correction for the H{\sc i} self-absorption by following the previous analysis of the cold H{\sc i} gas without CO emission (see Section 3.3 of Paper II).

\subsection{X-rays}\label{subsection:xrays}
We used $Suzaku$ archive data of RX J1713.7$-$3946 taken from Data Archives and Transmission System (DARTS at ISAS/JAXA). The observations performed 15 pointings toward the main features and 2 OFF pointings of RX J1713.7$-$3946 and were published by \cite{takahashi2008} and \cite{tanaka2008} expect for the 4 pointings observed in 2010 February. Previous and current observations are summarized in Table \ref{tab1}, and the FoV of each observation is shown in Figure \ref{fig1}. Active detector systems aboard the $Suzaku$ satellite are the X-ray Imaging Spectrometer \citep[XIS;][]{koyama2007} and the Hard X-ray Detector \citep[HXD;][]{takahashi2007}. The XIS consists of four CCD cameras placed at the foci of X-ray Telescopes \citep[XRTs;][]{serlemitsos2007}. We analyzed only XIS data in the present paper. The spaced-row charge injection \citep[SCI;][]{nakajima2008,uchiyama2009} was used in the latter 4 pointings (see also Table \ref{tab1}). Unfortunately, XIS 2 was on closed access since 2006 November 9, possibly owing to a micrometeorite damage. XIS 0 showed an anomaly in Segment A on 2009 June 23. Thus, for the data in the latter 4 pointings we used XIS 0 (expect for Segment A), XIS 1 and XIS 3. We used ``cleaned event files'' processed and screened by versions 2.0 or 2.4 $Suzaku$ pipeline depending on observation dates. First, we created the photon count images from the cleaned event files in the energy bands 1--5 keV and 5--10 keV. Here, we subtracted the non X-ray background (NXB) using \textbf{xisnxbgen}, which estimates NXB count rate based on night Earth observation data. Then, we corrected for XRT vignetting effects by simulating flat field images with \textbf{xissim} \citep{ishisaki2007}. Additionally, we masked the region of $^{55}$Fe calibration sources in the energy band 5--10 keV. Finally, we smoothed the images by using a Gaussian kernel with a FWHM of 45$\arcsec$. We performed data reduction with the version 6.11 of the HEAsoft tools.

\vspace*{0.25cm}
\section{Analysis}\label{section:analysis}

\subsection{Large-scale CO, H{\sc i} and X-ray distributions}\label{section:3.1}
Figure \ref{fig2} shows mosaic images of RX J1713.7$-$3946 which were constructed by using the data from XIS 0+1+2+3. Figures \ref{fig2}a and \ref{fig2}b show the soft band (1--5 keV) and hard band (5--10 keV) images, respectively. The unit for the images is 10$^{-4}$ counts s$^{-1}$ pixel$^{-1}$, and the pixel size is $\sim$16.7$\arcsec$. We find that the soft- and hard-band images are very similar with each other, which is already discussed in the previous study by \cite{tanaka2008}. Figure \ref{fig2} shows the western rim clearly as well as several peaks of $\sim$10$\times$10$^{-4}$ counts s$^{-1}$ pixel$^{-1}$ in the northern rim and inside the SNR. In the soft band image, thick white circles indicate locations of the two bright point-like sources toward the inner part of the SNR. The left one is associated with a Wolf-Rayet star CD$-$39 11212B \citep{pfeffermann1996}, which corresponds to two X-ray point sources cataloged (1WGA J1714.4$-$3945 and EXO 1710$-$396; see also Table \ref{tab2}). The other one is thought to be a neutron star because of its X-ray spectral characteristics \citep{lazendic2003}, which is cataloged as an X-ray point source (1WGA J1713.4$-$3949) and a pulsar (PSR J1713$-$3949). We also showed the modified color scale image in the energy band 1--5 keV, which enhances the regions of the low photon counts $\sim$7$\times$10$^{-4}$ counts s$^{-1}$ pixel$^{-1}$. In addition to the localized peaks in X-rays, we find diffuse X-ray emission is extended inside the SNR. In order to estimate the level of this background X-rays in the 1--5 keV X-ray image, we show two histograms of the X-ray counts in Figure \ref{fig13} (see Appendix B); one is for the whole region observed with $Suzaku$ (Figure \ref{fig1}) and the other for the nine circles of 6-arcmin diameter without significant peaks inside the SNR (Figure \ref{fig13}). In the histogram inside the SNR, we find a peak at $\sim$3.86$\times$10$^{-4}$ counts s$^{-1}$ pixel$^{-1}$ and identify this level as the background within the SNR. On the other hand, we consider that a primary peak at $\sim$1.16$\times$10$^{-4}$ counts s$^{-1}$ pixel$^{-1}$ for the whole region indicates the background level outside the SNR. In Figure \ref{fig2}c we plotted the positions of the X-ray point sources, pulsars and Wolf-Rayet stars (Table \ref{tab2}) in order to test whether the X-ray distribution is influenced by these point sources. We see no excess toward the point sources in Figure \ref{fig2}c except for the two bright point-like sources marked in Figure \ref{fig2}a, and consider that X-ray features inside the SNR are not due to the point sources but are intrinsic to the SNR.

Figure \ref{fig3} shows four overlays of the $^{12}$CO($J$=2--1) distribution and X-ray images in the two energy bands 1--5 keV (Figures \ref{fig3}a and \ref{fig3}b) and 5--10 keV (Figures \ref{fig3}c and \ref{fig3}d), respectively. A $V_\mathrm{LSR}$ range of CO from $-20.2$ to $-9.1$ km s$^{-1}$ is shown in Figures \ref{fig3}a and \ref{fig3}c and that from $-9.1$ to 1.8 km s$^{-1}$ in Figures \ref{fig3}b and \ref{fig3}d. These velocity ranges correspond to that of the interacting molecular gas \citep{fukui2012}. Figure \ref{fig3} indicates that CO and X-rays show a good correlation at a pc scale as already noted by \cite{moriguchi2005}. It is remarkable that most of the X-ray features are found toward CO clumps. The most outstanding X-rays are seen in the west of the shell, where the strongest CO emission is located (Figures \ref{fig3}b and \ref{fig3}d), and the second brightest X-rays are in the north of the shell where CO emission is also distributed (Figures \ref{fig3}a and \ref{fig3}c). The southern part of the CO emission appears to delineate the southern rim of the SNR (Figures \ref{fig3}b and \ref{fig3}d), while the eastern shell with weak X-rays has only a few small CO features (Figures \ref{fig3}b and \ref{fig3}d).

The CO distribution is highly clumpy. In order to make a detailed comparison with the X-rays, we cataloged CO clumps in the $^{12}$CO($J$=1--0) data in Figure \ref{fig12} in Appendix A. We identified 22 CO clumps in total, which are selected by the following two criteria in the $^{12}$CO($J$=1--0) data; (1) the peak position is located within SNR boundary, (2) the peak brightness temperature is higher than 1 K, and (3) the total clump surface area defined as the region surrounded by the contour at half of the maximum integrated intensity is larger than a 3-beam area. We give their observed parameters in $^{12}$CO($J$=1--0) and $^{12}$CO($J$=2--1) in Table \ref{tab2}. 13 of them are identified either by \cite{fukui2003} or \cite{moriguchi2005}. The rest of the clumps are newly identified in the present work. Most of the CO clumps have a single velocity component of line width $\sim$ 3--5 km s$^{-1}$. Only clump O \citep{moriguchi2005} has two velocity components of $\sim$7.5 km s$^{-1}$ separation, and is divided into two clumps O and O{\tiny{b}}. Five of them {D{\tiny{W}}, G{\tiny{E}}, O{\tiny{b}}, O{\tiny{SW}} and Z} have molecular mass higher than 50 $M_{\odot}$ (see also Table \ref{tab2}), and four of them {C{\tiny{E}}, Q{\tiny{W}}, Z{\tiny{NW}} and Z{\tiny{NE}}} have molecular mass less than 50 $M_{\odot}$. We focus hereafter on the 18 CO clumps which have molecular mass greater than 50 $M_{\odot}$ as shown in Figure \ref{fig3}, so that derivation of the physical parameters is ensured for a quantitative comparison with the X-rays (sub-section \ref{subsection:xrays}). Except for the five clumps C, I, L, O{\tiny{SW}} and Z, which are located inside of the SNR boundary, most of the CO clumps (A, B, D, D{\tiny{W}}, E, G, G{\tiny{E}}, O, O{\tiny{b}}, Q, R, U and W) are distributed on the outer boundary of the SNR shell.

Finally, we compare the cold H{\sc i} gas without CO with the X-rays in the southeast-rim of the SNR (hereafter SE-rim; Paper II). The cold H{\sc i} gas has density around 100 cm$^{-3}$ and is likely interacting with the shock in a similar way to CO. Figure \ref{fig4} shows an enlarged view in the SE-rim overlaid with the H{\sc i} proton column density contours. The integration range is $-20.0$--$-11.0$ km s$^{-1}$. We apply the H{\sc i} self-absorption by following the analysis in Paper II. The lowest contour level and the contour interval in the H{\sc i} proton column density are 2.0 $\times$ 10$^{21}$ cm$^{-2}$ and 0.1 $\times$ 10$^{21}$ cm$^{-2}$, respectively. It is remarkable that the H{\sc i} distribution corrected for the self-absorption is complementary to the X-ray peaks in the low-photon-count region in the SE-rim.

\subsection{Detailed Comparison with the X-Rays}\label{section:3.2}

\begin{figure*}
\begin{center}
\includegraphics[width=182mm,clip]{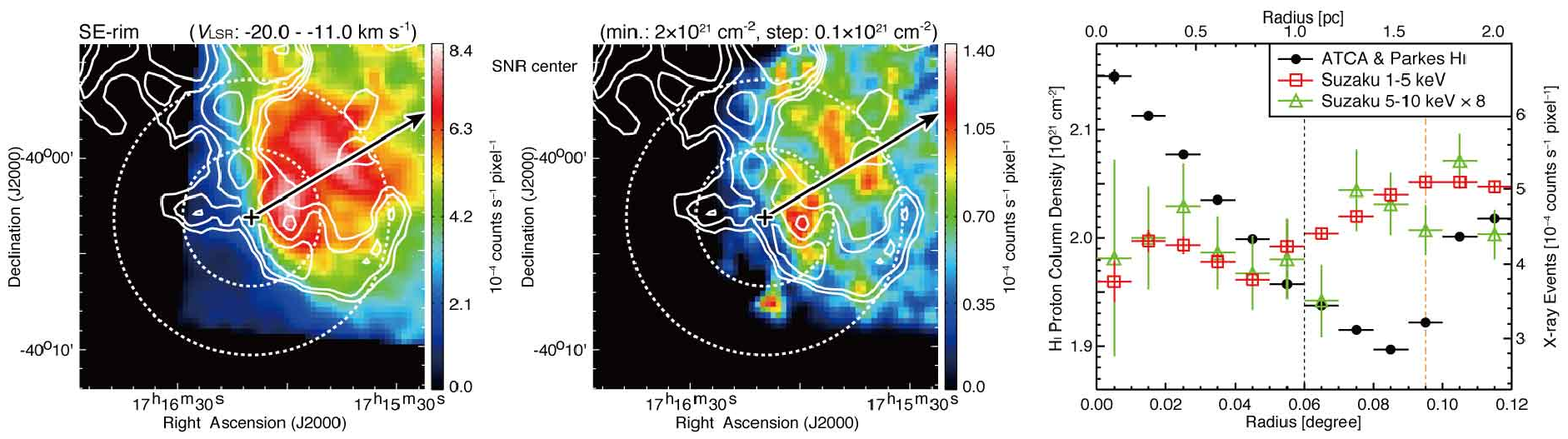}
\caption{Distribution of the cold H{\sc i} column density ($white$ $contours$) superposed on the $Suzaku$ images in the 1--5 keV (right) and 5--10 keV (middle) bands. Velocity range of integration and contour levels are shown in the top of the left and middle panels, respectively. Each arrow indicates the direction of the center of the SNR passing through the X-ray peak in the 1--5keV image. The crosses show the position which represents the cold H{\sc i} clump and is determined as the crossing point of the H{\sc i} cloud with the arrow, ($\alpha_{\mathrm{J2000}}$, $\delta_{\mathrm{J2000}}$) = ($17^{\mathrm{h}}$ $16^{\mathrm{m}}$ $9.3^{\mathrm{s}}$, $-40^{\circ}$ 3$\arcmin$ $10.6\arcsec$) (see also the text). The dashed white circles represent radii of 0$\fdg$06 and 0$\fdg$12 centered on the crosses. Right panel shows the radial profile around the crosses in H{\sc i} column density and the X-rays counts in the two energy bands (1--5 keV and 5--10 keV in Figure \ref{fig2}. The radial profile in the 5--10 keV band is scaled by a factor of 8 of the 1--5 keV band for the sake of direct comparison. The orange dash-dotted line indicates X-ray peak radius in the 1--5 keV band.}
\label{fig7}
\end{center}
\end{figure*}

\begin{figure*}
\begin{center}
\includegraphics[width=180mm,clip]{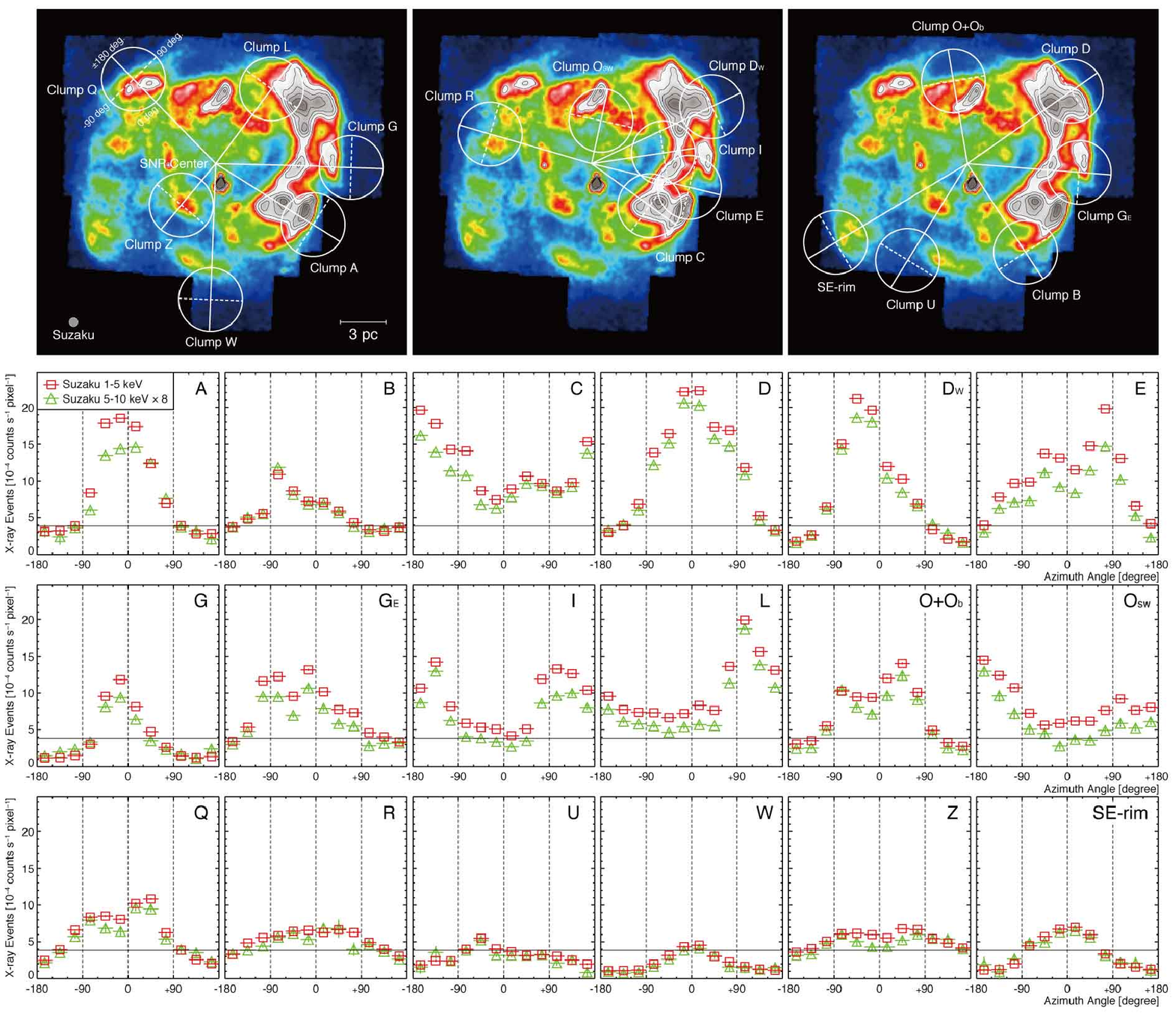}
\caption{$top$ $panels$: Same XIS mosaic image (1--5 keV) as Figure \ref{fig2} (c). Overlaid white circles are centered on the center of gravity of each CO/H{\sc i} clump. The azimuthal distribution of X-rays is averaged within each circle (radius = 0$\fdg$12). Each white solid straight line connects the centers of the circle and the SNR, and each dashed white line in the circle is vertical to the solid line. The azimuthal angle is measured from the solid line as the origin counterclockwise from $-$180 degrees to 180 degrees (e.g., clump Q in $top$ $panel$). $other$ $panels$: Azimuthal distributions of $Suzaku$ XIS 1--5 keV (red square) and 5--10 keV (green triangles are scaled by a factor of 8) averaged in the circles for each clump. The horizontal solid lines indicate the background level of the X-rays estimated inside the SNR for the energy band 1--5 keV (see the text).}
\label{fig8}
\end{center}
\end{figure*}

We here make a detailed comparison of spatial distributions between CO/H{\sc i} clumps and X-rays in images, radial and azimuthal distributions (Figures \ref{fig5}, \ref{fig7} and \ref{fig8}).

In Figure 5 left and middle panels of each row, we show the images of the CO integrated intensity overlayed on the distributions of the soft (1--5 keV, left) and hard (5--10 keV, middle) X-rays. The crosses in each image indicate the center of gravity of the CO clumps listed in Table \ref{tab4}, which is somewhat different from the peak position in Table \ref{tab2}, and each arrow indicates the direction of the center of the SNR. The dashed white circles represent radii 0$\fdg$06 and 0$\fdg$12 of the center of gravity. We see a trend that the X-rays are enhanced toward the CO, while the CO peak generally shows offsets from the X-ray peak.

In Figure \ref{fig5} right panels, we plot radial profiles of the CO integrated intensity and X-ray counts averaged at each radius for the 1--5 keV and 5--10 keV bands. In order to characterize quantitatively the radial distribution, we first identify the peak in the 1--5 keV radial distribution. In addition, we defined the separation from the center of gravity of CO clump to the X-ray peak in the radial distribution. 10 of the 17 clumps have peaks of the X-rays and positive X-ray slopes inside the peak in Figure \ref{fig5}. Clump W has no peak but also shows a clear positive X-ray slope. On the other hand, four of the CO clumps show negative slope and the other one show nearly flat slope. The radial distributions of the X-rays are generally smooth and monotonic so that linear approximation is reasonable in estimating the intensity gradients. Only the CO clumps O+O{\tiny{b}} show a complicated non-monotonic radial distribution of the X-rays, which may be due to blending of the two velocity components. We made least-squares fit to the X-rays by a straight line for simplicity for the 10 clumps within the peak, and for clump W and the other six clumps within a radius of 0.06 degrees. The values of the slope are listed in Table \ref{tab5}, and are shown as a histogram in Figure \ref{fig6}. 12 of the 16 clumps (75$\%$) show positive slopes in the X-rays, indicating that the X-rays show decrease toward the CO clump and are brightened in the surroundings of the CO clump. For the remaining 4 with negative or flat slopes except for clump Z, we also find clear relative enhancement of the X-rays in the surroundings of the CO clumps (in Figure \ref{fig5} left and middle panels). Clump Z has no clear X-ray depression toward the center, while enhanced X-rays are seen in its surroundings (Figure \ref{fig5} left and middle panels in the last low). We conclude that the enhanced X-rays around CO clumps are a general trend among the CO clumps, as is consistent with the previous result on clump C (Paper I). We note that the averaged behavior is that the CO clumps have a radius of 0.04 $\pm$ 0.01 degrees and the X-rays are distributed with a separation 0.07 $\pm$ 0.03 degrees from the center of each clump (see also Table \ref{tab4}).

In addition, we made a similar analysis on the cold H{\sc i} in the SE-rim. The H{\sc i} distribution estimated from self-absorption well delineates the outer boundary of the X-rays (Figure \ref{fig7}), and their relative distributions are similar to the case of clump A. The H{\sc i} column density distribution is fairly flat having no clear peak. Instead of using the H{\sc i} peak, we here draw a line passing through the center of the SNR and the soft X-ray peak, and define the H{\sc i} column density peak on this line. The radial distributions are plotted centered on this H{\sc i} peak. We find a trend similar to the CO clumps that the X-rays are clearly enhanced around the cold H{\sc i}.

\begin{deluxetable*}{lccccccccc}
\tablecaption{Results of Radial and Azimuthal Distribution}
\tablewidth{0pt}
\tablehead{\multicolumn{1}{c}{Name} & $\alpha_{\mathrm{J2000}}$ & $\delta_{\mathrm{J2000}}$ & $V_{\mathrm{LSR}}$ & Radius & \scalebox{0.8}[1]{Separation} & Peak Intensity & Angle (Fraction) & \scalebox{0.8}[1]{Interacting Mass} \\
 & ($^{\mathrm{h}}$ $^{\mathrm{m}}$ $^{\mathrm{s}}$) & ($^{\circ}$ $\arcmin$ $\arcsec$) & (km $\mathrm{s^{-1}}$) & (degree) & (degree) & \scalebox{0.6}[1]{($\times$10$^{-4}$ counts s$^{-1}$ pixel)} & (degree), ($\%$) & ($M_{\odot}$) \\
\multicolumn{1}{c}{(1)} & (2) & (3) & (4) & (5) & (6) & (7) & (8) & (9)}
\startdata
A................... & 17 11 39.1 & $-39$ 59 22.2 & $-$14.5--\phantom{0}$-$5.7 & 0.050 & 0.065 & 18.50$\pm$0.07 & $-$120--$+$120, (\phantom{0}67) & 460$\pm$60 \\
B................... & 17 12 25.3 & $-40$ 05 37.5 & $-$12.7--\phantom{0}$-$3.6 & 0.050 & 0.025 & 10.88$\pm$0.05 & $-$150--\phantom{0}$+$90, (\phantom{0}67) &130$\pm$20 \\
C................... & 17 12 25.3 & $-39$ 55 07.4 & $-$16.6--\phantom{0}$-$7.4 & 0.040 & 0.035 & 19.64$\pm$0.07& $-$180--$+$180, (100) & 400$\pm$30 \\
D................... & 17 11 31.9 & $-39$ 29 13.6 & $-$14.2--\phantom{0}$-$4.6 & 0.040 & \nodata & 22.25$\pm$0.07 & $-$150--$+$150, (\phantom{0}83) & 240$\pm$20 \\
D{\tiny{\textsc{W}}}................ & 17 11 12.8 & $-39$ 32 43.5 & \phantom{0}$-$7.2--\phantom{0}\phantom{0}4.8 & 0.040 & \nodata & 21.18$\pm$0.07 & $-$120--\phantom{0}$+$90, (\phantom{0}58) &  \phantom{0}80$\pm$11  \\
E................... & 17 11 38.7 & $-39$ 50 56.8 & \phantom{0}$-$8.1--\phantom{0}$-$4.0 & 0.040 & 0.105 & 19.81$\pm$0.07 & $-$180--$+$180, (100) & 159$\pm$13 \\
G................... & 17 10 54.5 & $-39$ 46 25.7 & $-$16.4--\phantom{0}$-$6.8 & 0.050 & 0.075 & 11.81$\pm$0.09 & \phantom{0}$-$60--\phantom{0}$+$60, (\phantom{0}33) &100$\pm$30\\
G{\tiny{\textsc{E}}}................. & 17 11 22.2 & $-39$ 47 34.0 & $-$15.1--\phantom{0}$-$9.8 & 0.050 & 0.025 & 13.16$\pm$0.13 & $-$150--$+$150, (\phantom{0}83) & 140$\pm$14 \\
I.................... & 17 12 09.3 & $-39$ 43 11.9 & $-$16.4--\phantom{0}$-$4.6 & 0.030 & 0.075 & 14.22$\pm$0.07 & $-$180--$+$180, (100) & 103$\pm$\phantom{0}9 \\
L................... & 17 12 28.2 & $-39$ 28 45.9 & $-$17.7--\phantom{0}$-$5.7 & 0.045 & 0.105 & 19.91$\pm$0.07 & $-$180--$+$180, (100) & 370$\pm$30 \\
O................... & 17 13 48.9 & $-39$ 26 24.9 & $-$11.3--\phantom{0}\phantom{0}$-$1.6 & 0.055 & 0.095 & 14.02$\pm$0.14 &$-$120--$+$120, (\phantom{0}67) &\phantom{0}41$\pm$\phantom{0}5 \\
O{\tiny{b}}................. & 17 13 46.9 & $-39$ 26 45.9 & \phantom{0}$-$4.6--\phantom{0}\phantom{0}\phantom{0}2.2 & 0.035 & 0.085 & 13.92$\pm$0.14 &$-$120--$+$120, (\phantom{0}67) &\phantom{0}53$\pm$\phantom{0}7\\
O{\tiny{\textsc{SW}}}............... & 17 13 22.4 & $-39$ 35 50.6 & \phantom{0}$-$2.7--\phantom{0}\phantom{0}\phantom{0}0.1 & 0.030 & 0.085 & 14.5\phantom{0}$\pm$0.2\phantom{0} & $-$180--$+$180, (100) & \phantom{0}60$\pm$\phantom{0}5  \\
Q................... & 17 15 09.6 & $-39$ 38 48.7 & \phantom{0}$-$5.1--\phantom{0}\phantom{0}\phantom{0}0.3 & 0.040 & 0.025 & 10.83$\pm$0.12 & $-$150--$+$120, (\phantom{0}75) &\phantom{0}81$\pm$\phantom{0}9 \\
R................... & 17 15 32.4 & $-39$ 39 28.5 & \phantom{0}$-$5.3--\phantom{0}$-$1.2 & 0.035  & 0.025 & \phantom{0}6.7\phantom{0}$\pm$0.2\phantom{0}& $-$150--$+$150, (\phantom{0}83) &\phantom{0}56$\pm$\phantom{0}6 \\
U................... & 17 14 44.5 & $-40$ 07 24.4 & \phantom{0}$-$6.1--\phantom{0}$-$3.4 & 0.030  & 0.105 & \phantom{0}5.54$\pm$0.07 & \phantom{0}$-$90-- \phantom{0}\phantom{0}\phantom{0}0, (\phantom{0}25) &\phantom{0}15$\pm$\phantom{0}5 \\
W.................. & 17 13 40.2 & $-40$ 16 16.9 & \phantom{0}$-$8.3--\phantom{0}$-$1.6 & 0.080  & \nodata & \phantom{0}4.58$\pm$0.07 & \phantom{0}$-$30--\phantom{0}$+$30, (\phantom{0}17) &\phantom{0}70$\pm$30 \\
Z................... & 17 14 14.3 & $-39$ 54 55.7 & $-$22.8--$-$17.5 & 0.025  & \nodata & \phantom{0}3.60$\pm$0.06 & $-$150--$+$180, (\phantom{0}92) &\phantom{0}66$\pm$\phantom{0}6 \\
H{\sc i} SE-rim..... & 17 16 09.3 & $-40$ 03 10.6 & $-$20.0--$-$11.0 & \nodata  & \nodata & \phantom{0}6.97$\pm$0.09 & \phantom{0}$-$90--\phantom{0}$+$60, (\phantom{0}42) &\phantom{0}56$\pm$11 \\
\cline{1-9}
O+O{\tiny{b}}........... & 17 13 49.7 & $-39$ 26 54.9 & \phantom{0}$-$9.8--\phantom{0}\phantom{0}\phantom{0}4.5 & 0.040 & 0.095 & 14.03$\pm$0.14 &$-$120--$+$120, (\phantom{0}67) &\phantom{0}94$\pm$\phantom{0}8\\
\cline{1-9}
D+D{\tiny{\textsc{W}}}.......... & \nodata & \nodata & \nodata & \nodata & \nodata & 21.72$\pm$0.05 & \nodata & 320$\pm$30\\
O+O{\tiny{b}}+O{\tiny{\textsc{SW}}}. & \nodata & \nodata & \nodata & \nodata & \nodata & 14.19$\pm$0.09 & \nodata & 154$\pm$10 \\
G+G{\tiny{\textsc{E}}}........... & \nodata & \nodata & \nodata & \nodata & \nodata & 12.49$\pm$0.08 & \nodata & 240$\pm$30 \\
E+I............... & \nodata & \nodata & \nodata & \nodata & \nodata & 17.02$\pm$0.05 & \nodata & 260$\pm$20
\enddata
\label{tab4}
\tablecomments{Col. (1): Clump name. Cols. (2)--(3): Position of the center of gravity with the $^{12}$CO($J$=2--1) integrated intensity (except for the SE-rim). Cols. (4): Integration range in velocity for the estimation of  the $^{12}$CO($J$=2--1) or H{\sc i}. (5): Radius of a CO clump defined as the radial distance from the center of gravity of a CO clump to the point where the intensity is at half maximum. (6) Separation of the X-ray peak from the center of gravity of each CO/H{\sc i} clump. (7) X-ray peak intensity with the statistical error around each clump shown in Figure \ref{fig8}. (8) Azimuth angle range of the X-rays (1--5 keV) above the background level estimated in the Appendix A (see also Section \ref{section:3.1}). (9) Interacting clump mass defined as the total CO/H{\sc i} mass within the azimuth angle range of the X-rays (1--5 keV) for each clump.\vspace*{0.2cm}}
\end{deluxetable*}

\begin{figure}
\vspace*{1.4cm}
\begin{center}
\includegraphics[width=86mm,clip]{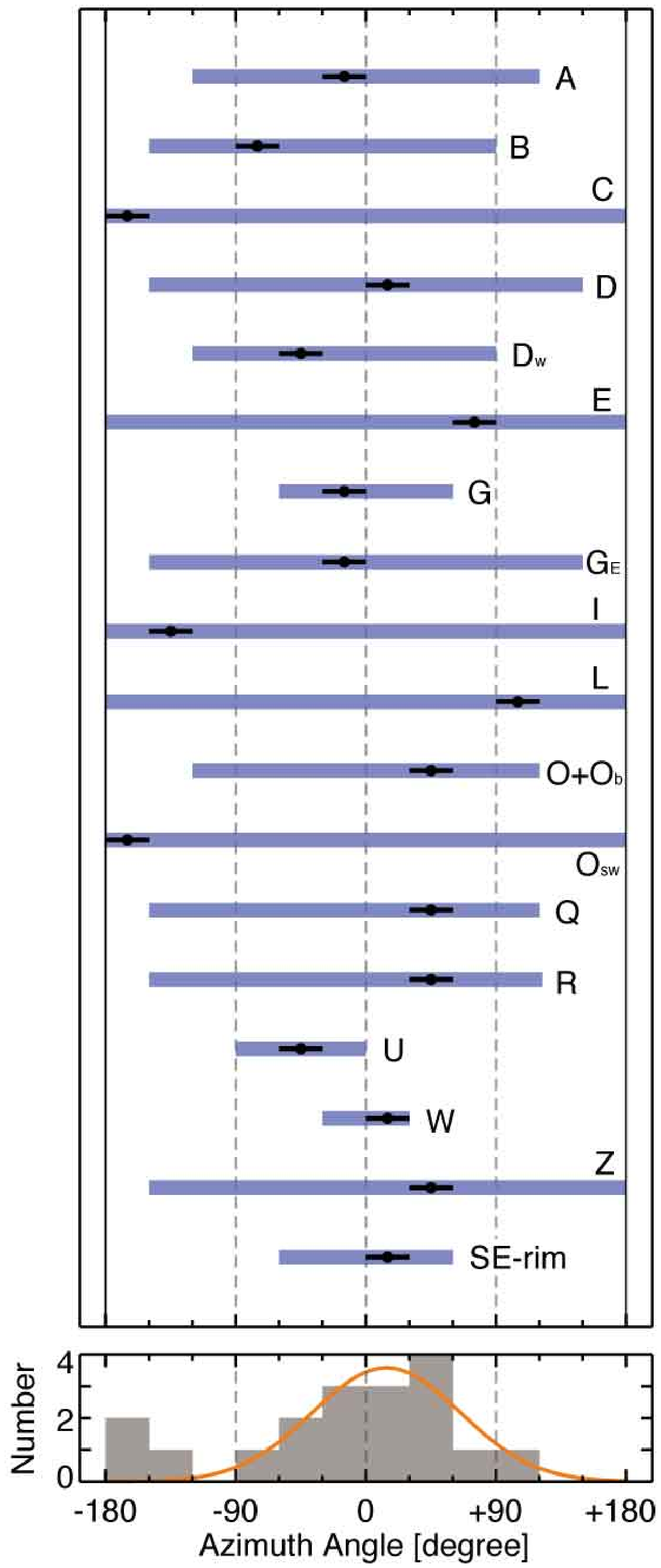}
\caption{The azimuthal angular extent of the X-rays above the background level in each CO/H{\sc i} clump for the 1--5 keV band image ($top$ $panel$; see also Table 3). Each black dot indicates the angle of X-ray peak intensities in Figure \ref{fig8}. The histogram of the X-ray peak positions is shown in $lowest$ $panel$. The red curve indicates the fitting result by a Gaussian function for the clumps from $-$90 degrees to $+$120 degrees.}
\label{fig9}
\end{center}
\end{figure}

In Figure \ref{fig8} top three panels show the same XIS mosaic images (1--5 keV) as in Figure \ref{fig2}c. The lines connecting the CO/H{\sc i} clump positions (shown in Figures \ref{fig5}, \ref{fig7} and Table \ref{tab4}) and the center of the SNR ($l$, $b$) = (347$\fdg$3, $-0\fdg5$) or ($\alpha_{\mathrm{J2000}}$, $\delta_{\mathrm{J2000}}$) = ($17^{\mathrm{h}}$ $13^{\mathrm{m}}$ $34^{\mathrm{s}}$, $-39^{\circ}$ 48$\arcmin$ $17\arcsec$) are taken as the origins of the azimuth angle, which is measured counterclockwise. The azimuthal angular distribution of the X-rays around each CO/H{\sc i} clump is estimated with respect to the direction of the center of the SNR. We measured the azimuthal angular extent of the X-rays in each clump in the 1--5 keV band image by adopting the background level inside the SNR as the threshold (Section \ref{section:3.1}). The results are shown in Figure \ref{fig8} and Table \ref{tab4}. Figure \ref{fig9} upper panel shows the range of angles and Figure \ref{fig9} lower panel a histogram of the peak angle. The five clumps inside the SNR, clumps C, E, I, L and O{\tiny{SW}}, are fully surrounded by the X-rays, which show a peak toward the azimuthal angle $-$180--$-$120 degrees (clumps C, I and O{\tiny{SW}}) or $+$60--$+$120 degrees (clump E and L). These distributions indicate that the clumps inside of the SNR are surrounded by the enhanced X-rays, whereas those on the border of the SNR have enhanced X-rays only toward the center of the SNR. In Figure \ref{fig9} lower panel we see a trend that the X-rays are enhanced at an azimuthal angle around 0, while scattering is large, $\pm$60 degrees (clumps A, D, D{\tiny{W}}, G, G{\tiny{E}}, O+O{\tiny{b}}, Q, R, U, W, Z and SE-rim). The histogram was fitted by a Gaussian function from $-$120 to 120 degrees, and we find that the best-fit parameters of center and sigma are 14$\pm$5 degrees and 51$\pm$5 degrees, respectively.

\begin{deluxetable}{lccl}
\vspace*{0.3cm}
\tabletypesize{\scriptsize}
\tablecaption{Fitting Results of the X-rays Radial Distribution}
\tablewidth{0pt}
\tablehead{\multicolumn{1}{c}{Name} &\multicolumn{1}{c}{Fitting Range}& \multicolumn{1}{c}{Slope} &\multicolumn{1}{c}{Comments} \\
 & (degree) &\hspace*{-0.5cm}\scalebox{0.6}[1]{($\times$10$^{-4}$ counts s$^{-1}$ pixel degree$^{-1}$)}\hspace*{-1cm}& \\
\multicolumn{1}{c}{(1)} & (2) & (3) & \multicolumn{1}{c}{(4)}}
\startdata
A.......... & 0.00--0.07 &$+$152$\pm$\phantom{0}2 & \scalebox{0.9}[1]{positive slope with peak}\\
B.......... & 0.00--0.03 &\phantom{0}$+$40$\pm$10 & \scalebox{0.9}[1]{positive slope with peak}\\
C.......... & 0.00--0.04 &\phantom{0}$+$56$\pm$\phantom{0}6 & \scalebox{0.9}[1]{positive slope with peak}\\
D.......... & 0.00--0.06 &\phantom{0}$-$59$\pm$\phantom{0}2 & \scalebox{0.9}[1]{negative slope}\\
D{\tiny{\textsc{W}}}....... & 0.00--0.06 &\phantom{0}$-$84$\pm$\phantom{0}3 & \scalebox{0.9}[1]{negative slope}\\
E.......... & 0.00--0.11 &\phantom{0}$+$82$\pm$\phantom{0}1 & \scalebox{0.9}[1]{positive slope with peak}\\
G......... & 0.00--0.08 &\phantom{0}$+$57$\pm$\phantom{0}1 & \scalebox{0.9}[1]{positive slope with peak}\\
G{\tiny{\textsc{E}}}........ & 0.00--0.03 &\phantom{0}$+$40$\pm$20 & \scalebox{0.9}[1]{positive slope with peak}\\
I........... & 0.00--0.08 &\phantom{0}$+$70$\pm$\phantom{0}2 & \scalebox{0.9}[1]{positive slope with peak}\\
L.......... & 0.00--0.10 &\phantom{0}\phantom{0}$+$2$\pm$\phantom{0}1 & \scalebox{0.9}[1]{flat}\\
O+O{\tiny{b}}... & \nodata &\nodata & \nodata\\
O{\tiny{\textsc{SW}}}...... & 0.00--0.09 &\phantom{0}$+$43$\pm$\phantom{0}3 & \scalebox{0.9}[1]{positive slope with peak}\\
Q.......... & 0.00--0.03 &\phantom{0}$+$20$\pm$20 & \scalebox{0.9}[1]{positive slope with peak}\\
R.......... & 0.00--0.03 &\phantom{0}$-$10$\pm$10 & \scalebox{0.9}[1]{negative slope}\\
U.......... & 0.00--0.10 &\phantom{0}$+$19$\pm$\phantom{0}1 & \scalebox{0.9}[1]{positive slope with peak}\\
W......... & 0.00--0.06 &\phantom{0}$+$10$\pm$\phantom{0}2 & \scalebox{0.9}[1]{positive slope with no peak}\\
Z.......... & 0.00--0.06 &\phantom{0}$-$79$\pm$\phantom{0}4 & \scalebox{0.9}[1]{negative slope}
\enddata
\label{tab5}
\tablecomments{Col. (1): Clump name. Col. (2): Fitting range of X-rays in radial plot (see also the text for details). Col. (3): Slope of the fitted straight line of X-rays with 1 sigma error in the least-squares fitting. (4) Comments.\vspace*{0.2cm}}
\end{deluxetable}

We shall here test if the X-ray absorption by the ISM affects the X-ray images. The X-ray counts in 1--5 keV are generally about 8 times higher than in 5--10 keV and the distributions in the two bands are fairly similar with each other (Figures \ref{fig5} and \ref{fig7}). The ratio 8 is consistent with an X-ray photon index of $\sim$2.3 typical to the SNR \citep[e.g.,][]{tanaka2008}. Only clump L shows the largest difference from 8 by a factor of more than two in the plot, indicating a softer spectrum. The similarity of the two bands suggests that X-ray absorption is not significant, because the X-rays should show a much harder spectrum if absorption is significant. The optical depth due to the ISM at X-rays is proportional to the ($-$8/3)-th power of the energy and is expressed as follows \citep{Longair1994};
\begin{eqnarray}
\tau_{\mathrm{x}} = 2\times10^{-22}\phantom{0}N_{\mathrm{H}}\phantom{0}\mathrm{(cm^{-2})} \cdot \varepsilon^{-8/3} \phantom{0}\mathrm{(keV)},
\end{eqnarray}
where $N_{\mathrm{H}}$ (cm$^{-2}$) is the ISM column density and $\varepsilon$ (keV) is X-ray photon energy. The maximum ISM column density in the SNR is estimated to be 1$\times$10$^{22}$ cm$^{-2}$ toward the brightest CO peak, peak C \citep[e.g.,][]{sano2010}. Even for this column density absorption optical depths are 2, 0.3, 0.03 and 0.004 at 1 keV, 2 keV, 5 keV, and 10 keV, respectively, as calculated by equation (1), and absorption is not likely affect the X-ray distribution significantly. We made a test on the X-ray absorption toward clump C, where the X-rays show depression which may be possibly due to absorption. The X-ray intensity ratio toward clump C is $\sim$8.6 between the 1--5 keV and 5--10 keV bands as observed by $Suzaku$ (Sano et al. 2013b in prep.). By using Xspec we calculated the X-ray intensity integrated over the two energy bands for three different values of absorbing column density, i.e., 0.3, 1 and 3$\times$10$^{22}$ cm$^{-2}$, for an X-ray photon index of 2.3, and found that the intensity ratios between the two bands are $\sim$11.1, 7.0, and 3.5, respectively. Column density slightly less than 1$\times$10$^{22}$ cm$^{-2}$ fits the X-ray observations reasonably well as is consistent with the CO observations.

\vspace*{0.3cm}
\section{Discussion}\label{section:discussion}

\begin{figure*}
\begin{center}
\includegraphics[width=180mm,clip]{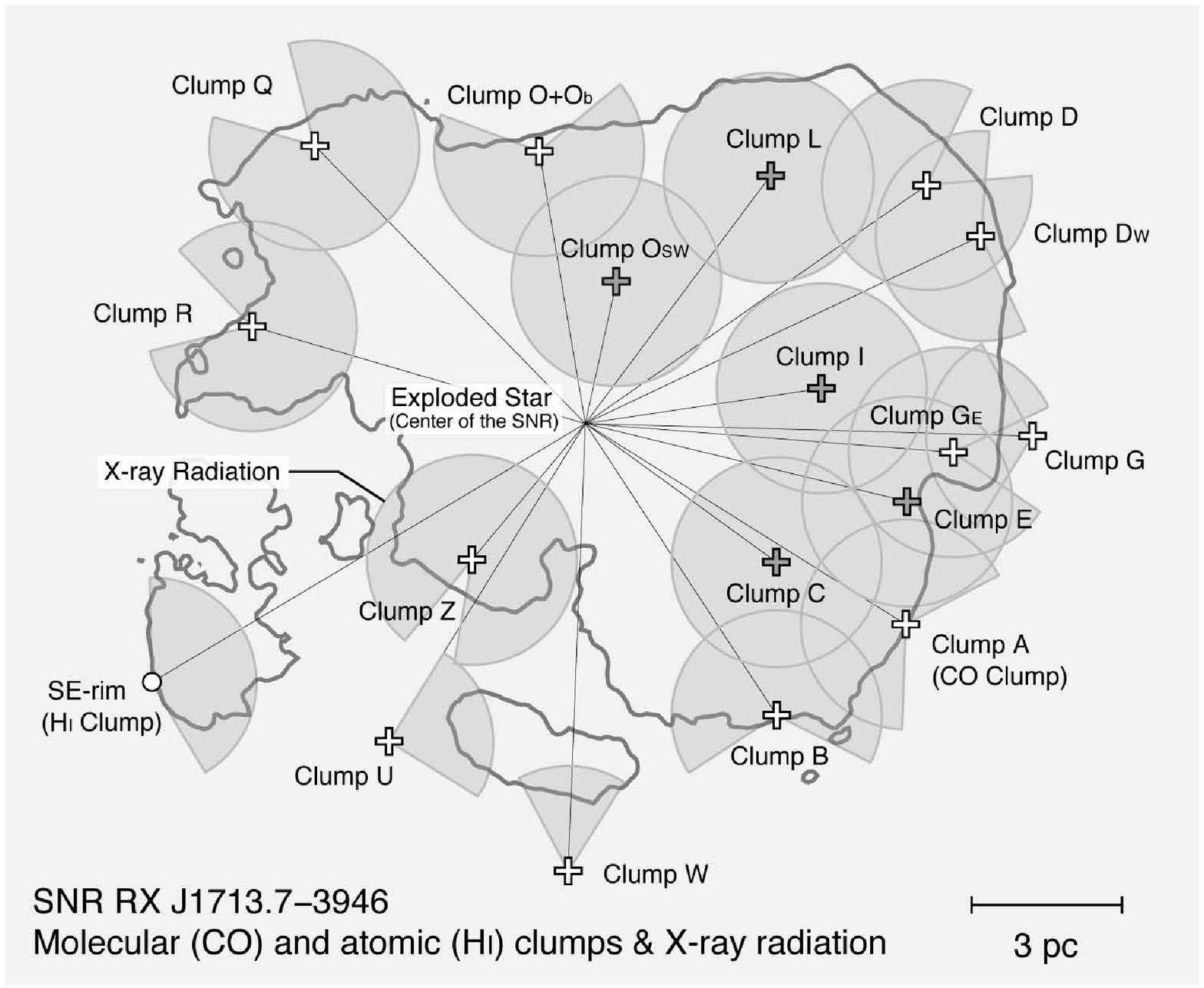}
\caption{Schematic image of the distribution of the molecular (CO) clumps (open crosses), atomic (H{\sc i}) clump (circle) and the X-rays (shaded partial or full circles) superposed on the $Suzaku$ 1--5 keV X-ray outer boundary of the SNR (gray contours). The black open crosses (clumps C, I, E, L, and O{\tiny{\textsc{SW}}}) indicate those fully surrounded by the X-rays.}
\label{fig10}
\end{center}
\end{figure*}%

Paper I showed that the synchrotron X-ray emission is enhanced around the CO clumps in the northwest region of RX J1713.7$-$3946. In the present paper, we have compared the X-rays and the CO and H{\sc i} dense clumps over the whole SNR, and present a schematic image of these results in Figure \ref{fig10}. This figure indicates that the CO and dense H{\sc i} clumps form an inhomogeneous shell and that the X-rays are enhanced around all the clumps. We infer that the five CO clumps C, E, I, L and O{\tiny{SW}} survived the SNR blast waves, being now embedded within the SNR, while the other twelve CO clumps and the H{\sc i} clump are shock-interacting on their inner side. The ISM shell was formed over a timescale of Myr by the stellar winds of an OB star which experienced an SN explosion 1600 yr ago (e.g., Papers I and II). Thus, the density in the cavity surrounded by the ISM shell is expected to be very low. The observational results also indicate that there is little dense gas left in the interior of the cavity (Figures \ref{fig3} and \ref{fig12}). According to the numerical simulations which studied the interaction of the ISM with the strong stellar winds from an O-type star \citep{weaver1977}, the gas density inside the evacuated wind bubble is $\sim0.01$ cm$^{-3}$, which applies to the interior of the cavity. The shock waves of the SNR first propagated in the stellar-wind cavity and then began interaction with the CO/H{\sc i} clumps some 1000 yrs ago as given by the ratio of the shell thickness and the shock velocity, 3 pc / 3000 km s$^{-1}$. According to MHD numerical simulations by \cite{inoue2012}, the CO/H{\sc i} clumps having density of $\sim10^2$--10$^3$ cm$^{-3}$ are surrounded by the interclump gas having density of $\sim$1 cm$^{-3}$ which is two orders of magnitude higher than that in the cavity. These authors show that the shock is stalled in the dense clumps. The shock velocity becomes $V_{\rm sh, clump} = V_{\rm sh, interclump} (n_{\rm interclump}/n_{\rm clump})^{0.5}$, where the interclump density $n_{\mathrm{interclump}}$ = 1 cm$^{-3}$ and the clump density $n_{\mathrm{clump}}$ = $10^2$--10$^3$ cm$^{-3}$. The shock velocity difference between the dense CO/H{\sc i} clumps and the interclump gas will become a factor of $\sim$10--30. They calculated that the temperature of the shocked dense gas becomes much lower than the temperature in the post-shock diffuse gas, and argued that the thermal X-ray emission from the CO/H{\sc i} clumps is strongly suppressed after the passage of the shock (see Section 4.3 of \citealp{inoue2012}). The interclump gas does not emit significant thermal X-rays either, because the density $\sim$1 cm$^{-3}$ is less than $\sim$2 cm$^{-3}$, the average density inside the SNR obtained from the upper limit of the thermal X-rays \citep{takahashi2008}.

An important consequence of the interaction is that the large velocity difference created between the clumps and the interclump space induces turbulence, which leads to turbulent dynamo action. The magnetic field is then amplified to as high as 1 mG, which is consistent with the field strength derived from rapid time variation of the X-ray filaments \citep{uchiyama2007}, while an alternative is that the fluctuations in the field orientation may explain the rapid time variation \citep{helder2012}. The synchrotron flux integrated in the line of sight is proportional to $B^{1.5}$ if the spectral index of electrons $p$ is 2.0 \citep[e.g.,][]{rybicki1979}. So, it is possible to enhance the X-ray radiation around the CO and H{\sc i} clumps. It is known that the power of the synchrotron X-ray emission is not enhanced by magnetic field amplification due to the effect of synchrotron cooling, if the amplification takes place in the vicinity of the forward shock where electrons are being accelerated \citep[e.g.,][]{nakamura2012}. In the present case, as discussed by \cite{inoue2009,inoue2010,inoue2012}, the magnetic field amplification owing to the shock-cloud interaction is effective at least 0.1 pc downstream of the shock front. This indicates that the synchrotron X-rays are emitted after the acceleration process, and thus the power of synchrotron X-ray is enhanced by the amplification. The observed power of the X-ray emission around the CO and H{\sc i} clumps is 2--7 times higher than the background level inside the SNR. Then, the magnetic field around the CO and H{\sc i} clumps is estimated to be 2--4 times higher than elsewhere in the SNR, if the X-ray enhancement is only due to the magnetic field amplification. The averaged magnetic field around the CO and H{\sc i} clumps becomes 30--60 $\mu$G if the initial field is assumed to be 15 $\mu$G \citep[e.g.,][]{tanaka2008}. The average field strength is also estimated by the width of synchrotron X-ray filaments as $\sim$100 $\mu$G \citep{bell2004, hiraga2005,ballet2006}. Note that the dependence of the synchrotron flux on the magnetic field strength can be much more sensitive than the above-mentioned standard case, because the high-energy electrons that contribute the X-ray synchrotron emission can be in the cut-off regime \citep{bykov2008}. Moreover, such enhanced magnetic field in turbulence may lead to more efficient acceleration than in the diffusive shock acceleration {\citep[][]{lazarian1999,hoshino2012}.

Finally, we discuss quantitative relationship between the CO/H{\sc i} interacting clump mass and the X-ray enhancement. First, we estimate the interacting clump mass (column (9) of Table \ref{tab4}) with the shock waves as defined by the total CO/H{\sc i} mass within the azimuth angle range of the X-rays with respect to the center of  gravity (column (8) of Table \ref{tab4}). For the case in which two CO clumps have small separation ($<$ 0.2 degrees) and the X-ray peak is situated between the CO clumps, we sum up the interacting clump masses and averaged the X-ray intensities (D$+$D{\tiny{W}}, O+O{\tiny{b}}+O{\tiny{\textsc{SW}}}, G+G{\tiny{\textsc{E}}} and E+I; see also Table \ref{tab4}). In Figure \ref{fig11}, we plot the CO/H{\sc i} interacting clump mass as a function the X-ray peak intensity. Here, we approximate the mass of the SE-rim to be 134 $M_{\odot}$ on the assumption that it has a large of 0.8 pc$^2$ along the X-ray boundary. The result, shown in Figure \ref{fig11}, indicates that the correlation between the interacting clump mass and the X-ray intensity is good with a correlation coefficient of $\sim$0.85 in double logarithm. We conclude that intensity is roughly proportional to the interacting mass of each CO/H{\sc i} clump at a pc scale. This result suggests that the ISM distribution is crucial in producing the non-thermal X-ray distribution in young SNRs.
\vspace*{0.1cm}
\begin{figure}
\begin{center}
\includegraphics[width=87mm,clip]{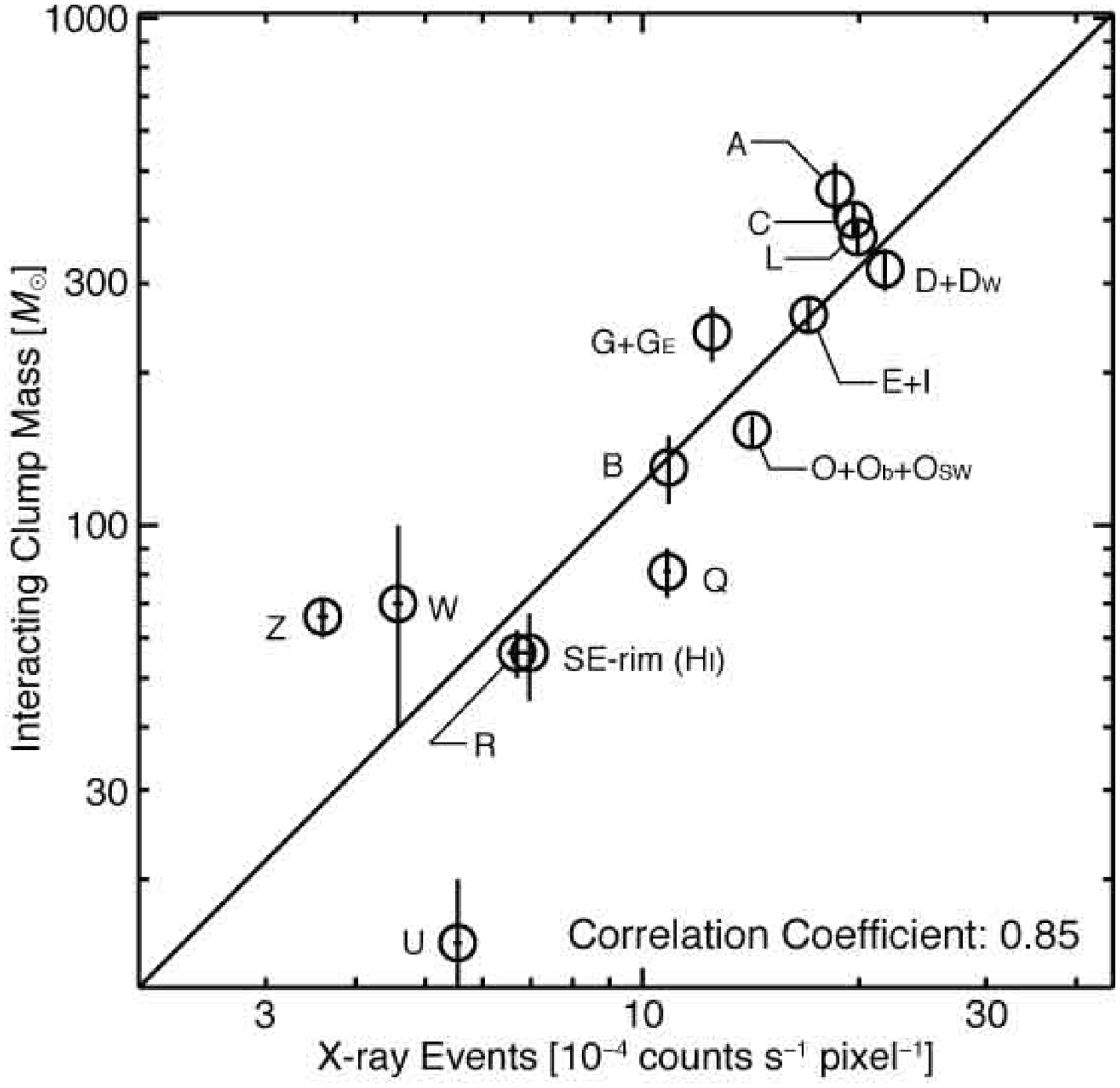}
\caption{Correlation plot between the X-ray peak intensity in the azimuthal distributions derived in Figure \ref{fig8} and the interacting clump mass with the shock waves, which is estimated by the CO or H{\sc i} mass within the azimuth angle range of the X-rays for each clump (see for more details the text, and column (9) of Table \ref{tab4}). The linear regression by the least-squares fitting is shown by the solid line, where the correlation coefficient is $\sim$0.85 in double logarithm.}
\label{fig11}
\end{center}
\vspace*{-0.2cm}
\end{figure}%

\section{Conclusions} \label{section:conclusions}
We summarize the present work as follows;

\begin{enumerate}
\item We have shown that all the major CO and H{\sc i} clumps with mass greater than 50 $M_{\odot}$ interacting with the shock waves in RX J1713.7$-$3946 are associated with the non-thermal X-rays. The X-rays are enhanced within $\sim$1 pc of the CO and H{\sc i} peaks, whereas at smaller scales down to 0.1 pc the CO peaks tend to be anti-correlated with the X-rays which are decreased toward the CO and H{\sc i} clumps. We have shown a good correlation between the CO/H{\sc i} clump mass interacting with the shock waves and the X-ray intensity.

\item The present findings in 1) are compared with numerical simulations of MHD in a realistic highly inhomogeneous density distribution by \cite{inoue2009, inoue2012}. These simulations indicate that the magnetic field is amplified around dense CO/H{\sc i} clumps as a result of enhanced turbulence induced by the shock-cloud interaction. We interpret that thus-amplified magnetic fields enhance the X-ray intensity, which depends on the 1.5-th power of the magnetic field strength. Such enhanced magnetic field may also lead to efficient acceleration additional to the DSA. More comparative studies of distribution of the X-rays and the ISM will allow us to have a deeper insight into the origin of the X-ray distribution.
\end{enumerate}

\acknowledgments
NANTEN2 is an international collaboration of 10 universities; Nagoya University, Osaka Prefecture University, University of Cologne, University of Bonn, Seoul National University, University of Chile, University of New South Wales, Macquarie University, University of Sydney, and University of ETH Zurich. This work was financially supported by a grant-in-aid for Scientific Research (KAKENHI, No. 21253003, No. 23403001, No. 22540250, No. 22244014, No. 23740149, No. 23740154 (T.I.), No. 22740119 and No. 24224005) from MEXT (the Ministry of Education, Culture, Sports, Science and Technology of Japan). This work was also financially supported by the Young Research Overseas Visits Program for Vitalizing Brain Circulation (R2211) and the Institutional Program for Young Researcher Overseas Visits (R29) by JSPS (Japan Society for the Promotion of Science) and by the Mitsubishi Foundation and by the grant-in-aid for Nagoya University Global COE Program, “Quest for Fundamental Principles in the Universe: From Particles to the Solar System and the Cosmos,” from MEXT. This research made use of data obtained from Data ARchives and Transmission System (DARTS), provided by Center for Science-satellite Operation and Data Archive (C-SODA) at ISAS/JAXA.

\section*{Appendix A\\Velocity Channel Distributions of the CO Clumps}
In order to clarify a relationship between X-rays and CO clumps, it is necessary to identify all the CO clumps interacting with the SNR blast waves. The previous studies \citep[e.g.,][]{fukui2003, moriguchi2005, sano2010} identified most of the CO clumps, but it was not complete because a velocity range was limited (e.g., $V_{\mathrm{LSR}}$: $-12$--$-3$ km s$^{-1}$ in \citealp{moriguchi2005}). We therefore identified all the CO clumps interacting with the SNR blast waves for a wide velocity range, such as in \cite{fukui2012} ($V_{\mathrm{LSR}}$: $-20$--$+2$ km s$^{-1}$). We used the velocity channel maps and appropriate criteria to identify other CO clumps (see also Section \ref{section:3.1}). Figure \ref{fig12} shows the velocity channel distribution of $^{12}$CO($J$=1--0, 2--1) every 2 km s$^{-1}$ from $-$24 km s$^{-1}$ to 8 km s$^{-1}$ superposed on the $Suzaku$ X-ray distribution (1--5 keV). We also plotted the position of each CO clump interacting with the SNR. The clumps, C{\tiny{E}}, D{\tiny{W}}, G{\tiny{E}}, O{\tiny{b}}, O{\tiny{SW}}, Q{\tiny{W}}, Z, Z{\tiny{NW}} and Z{\tiny{NE}}, are newly identified in present work.

\section*{Appendix B\\The background level of X-rays}
In order to estimate the azimuth distribution of the X-rays in Figure \ref{fig8}, we estimated the background level of the X-rays in the energy band 1--5 keV inside the SNR. Figure \ref{fig13} shows two histogram of the X-rays. One is that extracted from the whole area observed as shown in Figure \ref{fig1}, and the other the typical inner part of the SNR. The latter is extracted from nine circles of 6$\arcmin$ diameter which include no significant peaks of the X-rays. The central circle is at  ($\alpha_{\mathrm{J2000}}$, $\delta_{\mathrm{J2000}}$) = ($17^{\mathrm{h}}$ $13^{\mathrm{m}}$ $52.8^{\mathrm{s}}$, $-39^{\circ}$ 49$\arcmin$ $12.0\arcsec$) and the other eight have offsets $\pm \sim11 \arcmin$ from the center in $\alpha_{\mathrm{J2000}}$ and/or $\delta_{\mathrm{J2000}}$ as shown in Figure \ref{fig13}b. We consider that the peak of the histogram for the whole ($\sim$1.16 $\times$ 10$^{-4}$ counts s$^{-1}$ pixel) is the typical background level outside the SNR, and that the primary peak of the histogram  for the inner part ($\sim$3.86 $\times$ 10$^{-4}$ counts s$^{-1}$ pixel) indicates the background level of the SNR interior. We used the excess counts from the background level as those of the individual X-ray features in deriving the azimuthal distribution in Figure \ref{fig8}.

\begin{figure*}
\begin{center}
\figurenum{12}
\includegraphics[width=180mm,clip]{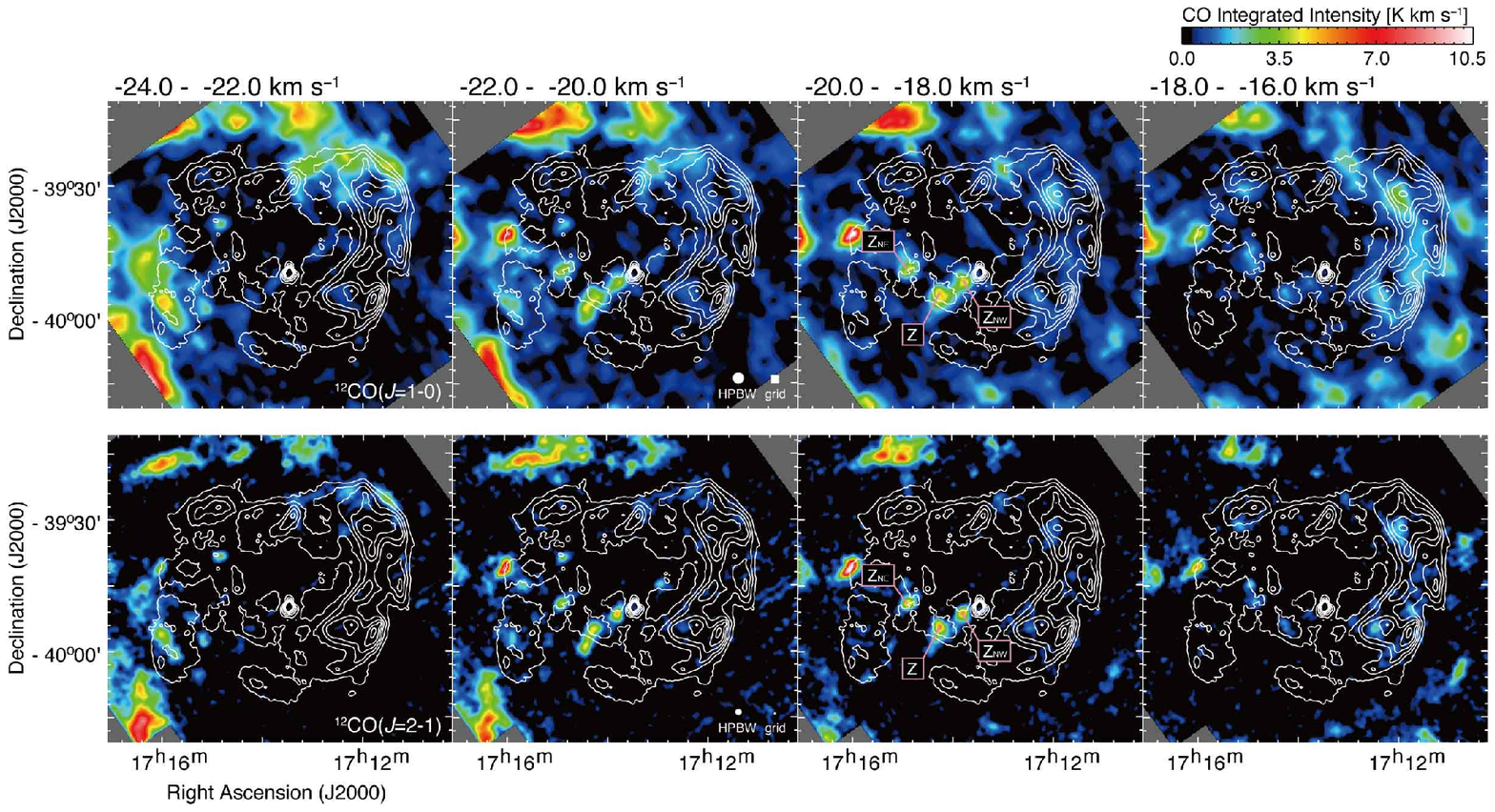}
\caption{Velocity channel distributions of the $^{12}$CO($J$=1--0) ($top$ $panels$ in false color) and $^{12}$CO($J$=2--1) ($bottom$ $panels$ in false color) emissions overlaid on the $Suzaku$ X-ray contours for the energy band 1--5 keV. Each panel of CO shows intensity distributions integrated every 2 km s$^{-1}$ in a velocity range from $-$24 to 8 km s$^{-1}$ following the color code shown on the upper right. In the X-ray distribution, the lowest contour level and the contour interval are 2.1 and 0.6 $\times$ 10$^{-4}$ counts s$^{-1}$ pixel$^{-1}$, respectively. The CO clumps shown in Table \ref{tab3} are also plotted.}
\label{fig12}
\end{center}
\end{figure*}%

\begin{figure*}
\begin{center}
\figurenum{12}
\includegraphics[width=180mm,clip]{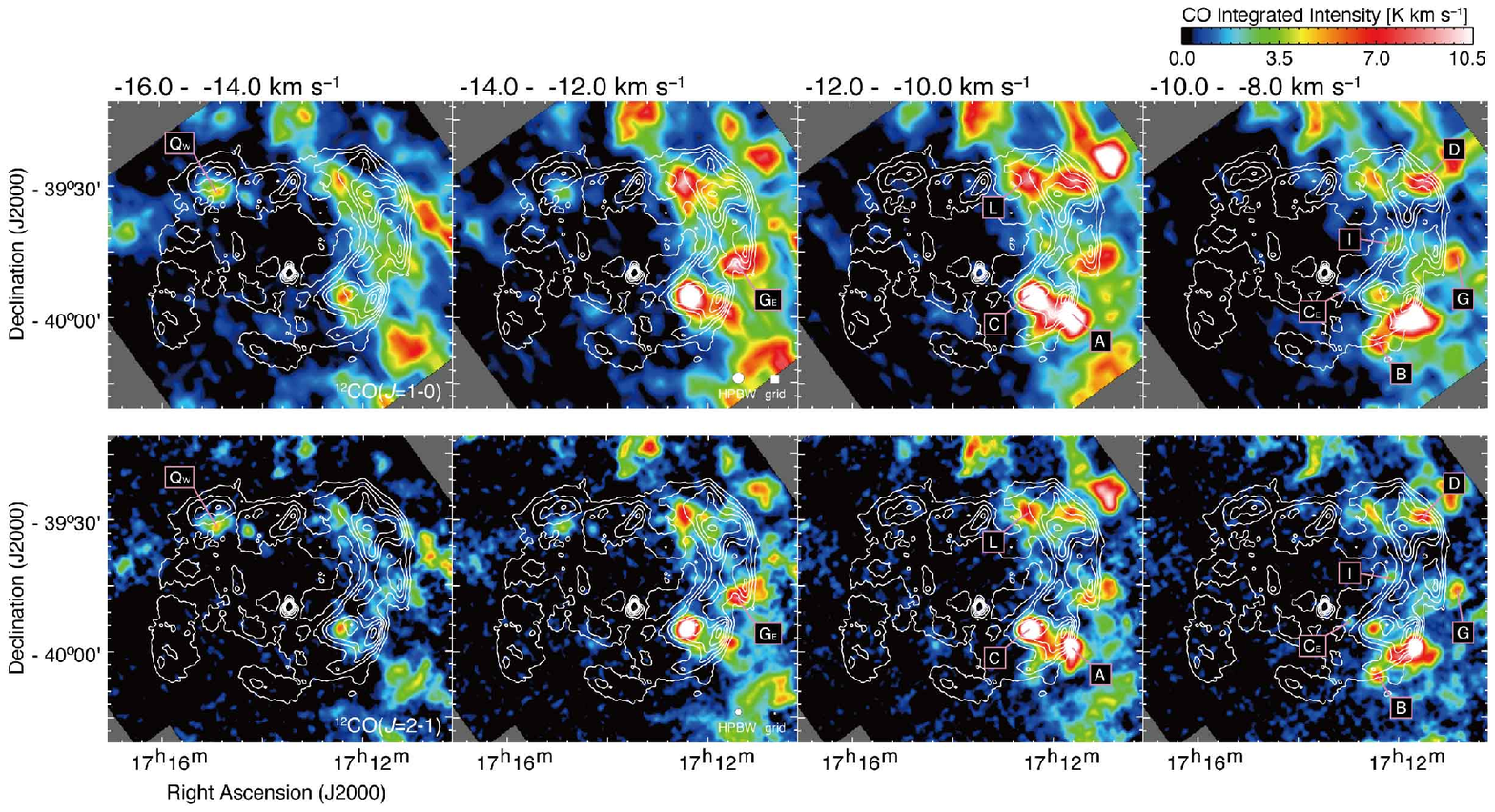}
\caption{$continued.$}
\label{fig12}
\end{center}
\end{figure*}%

\begin{figure*}
\begin{center}
\figurenum{12}
\includegraphics[width=180mm,clip]{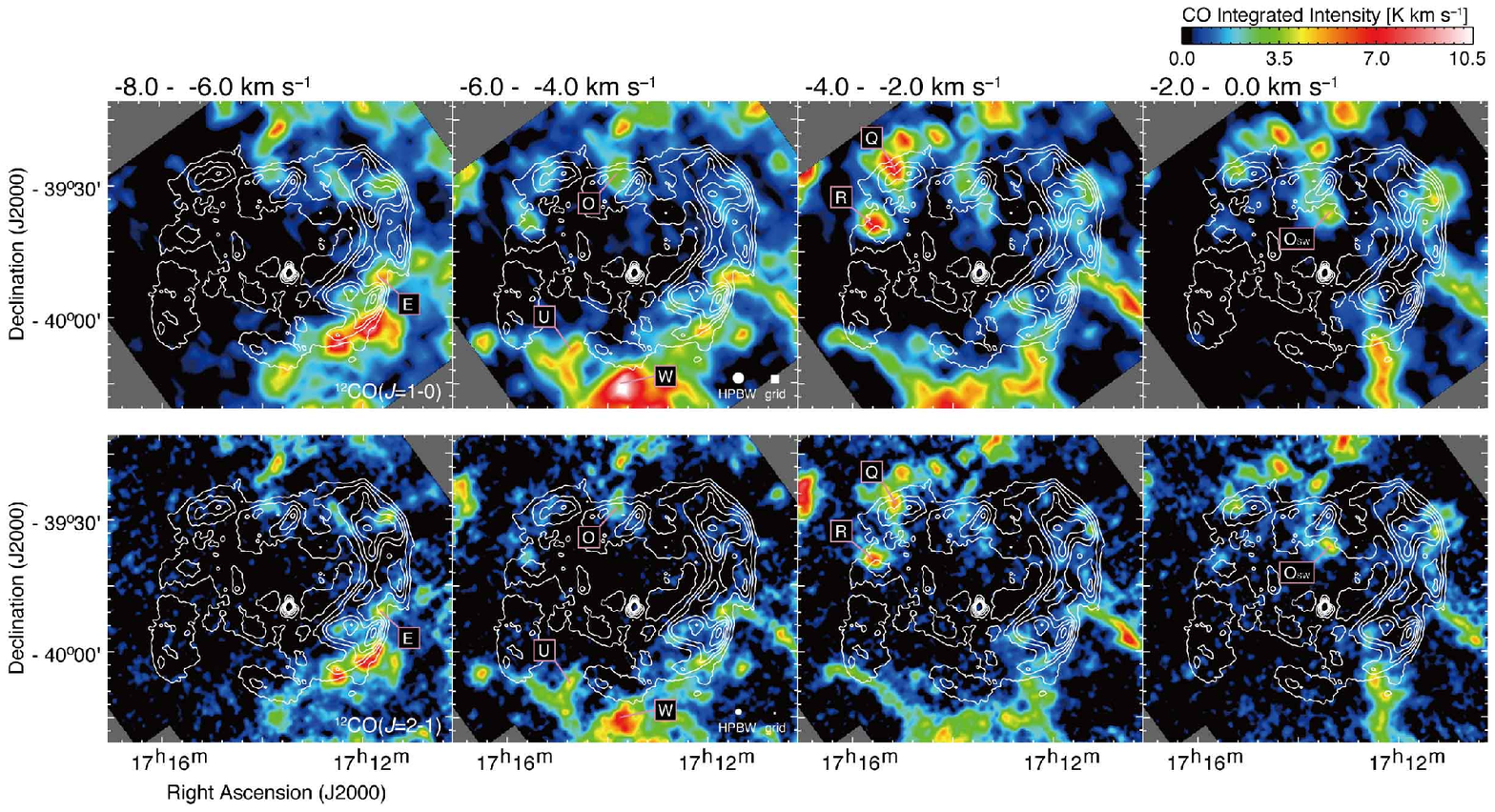}
\caption{$continued.$}
\label{fig12}
\end{center}
\end{figure*}%

\begin{figure*}
\begin{center}
\figurenum{12}
\includegraphics[width=180mm,clip]{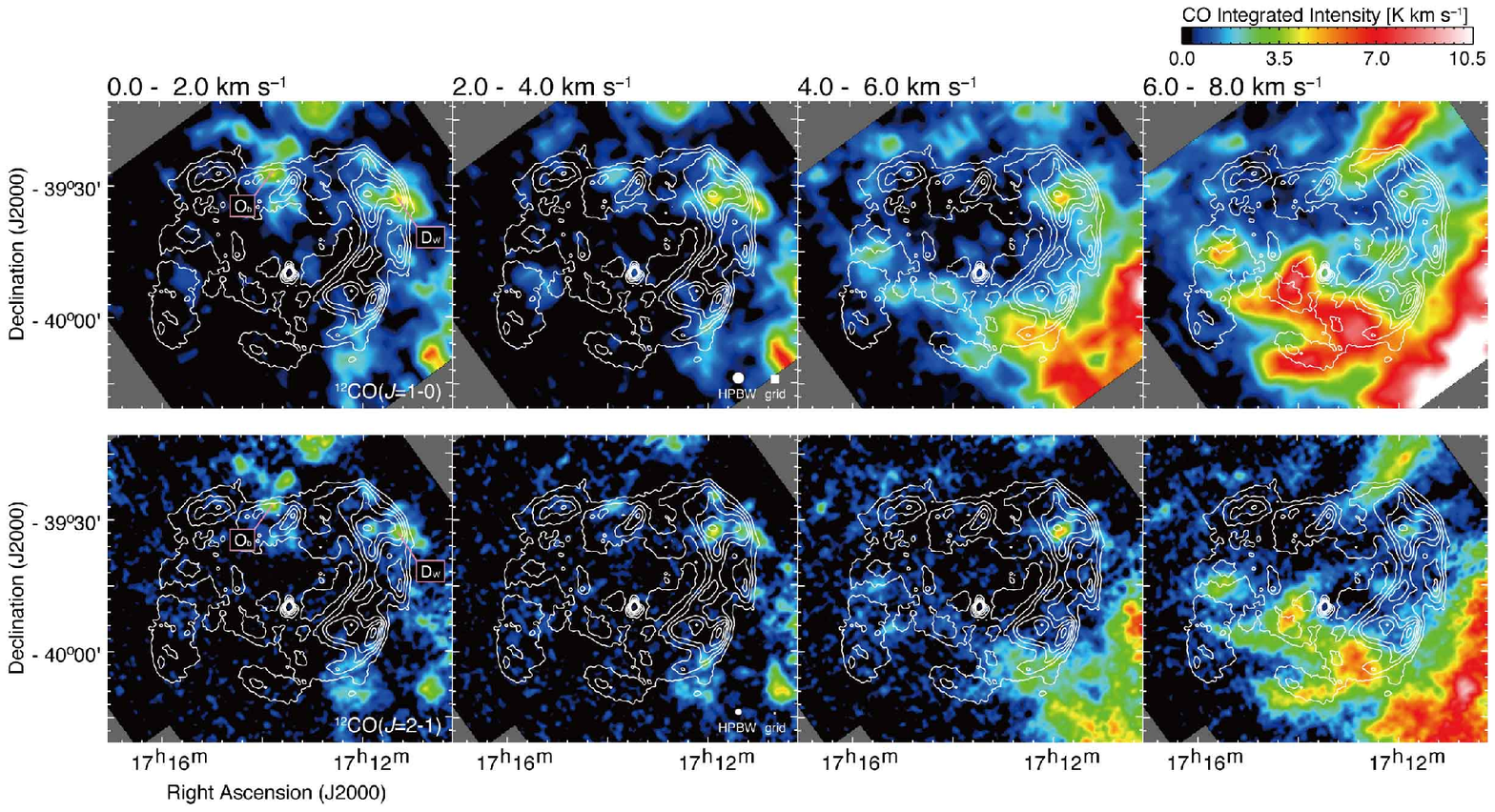}
\caption{$continued.$}
\label{fig12}
\end{center}
\end{figure*}%

\begin{figure*}
\begin{center}
\figurenum{13}
\includegraphics[width=180mm,clip]{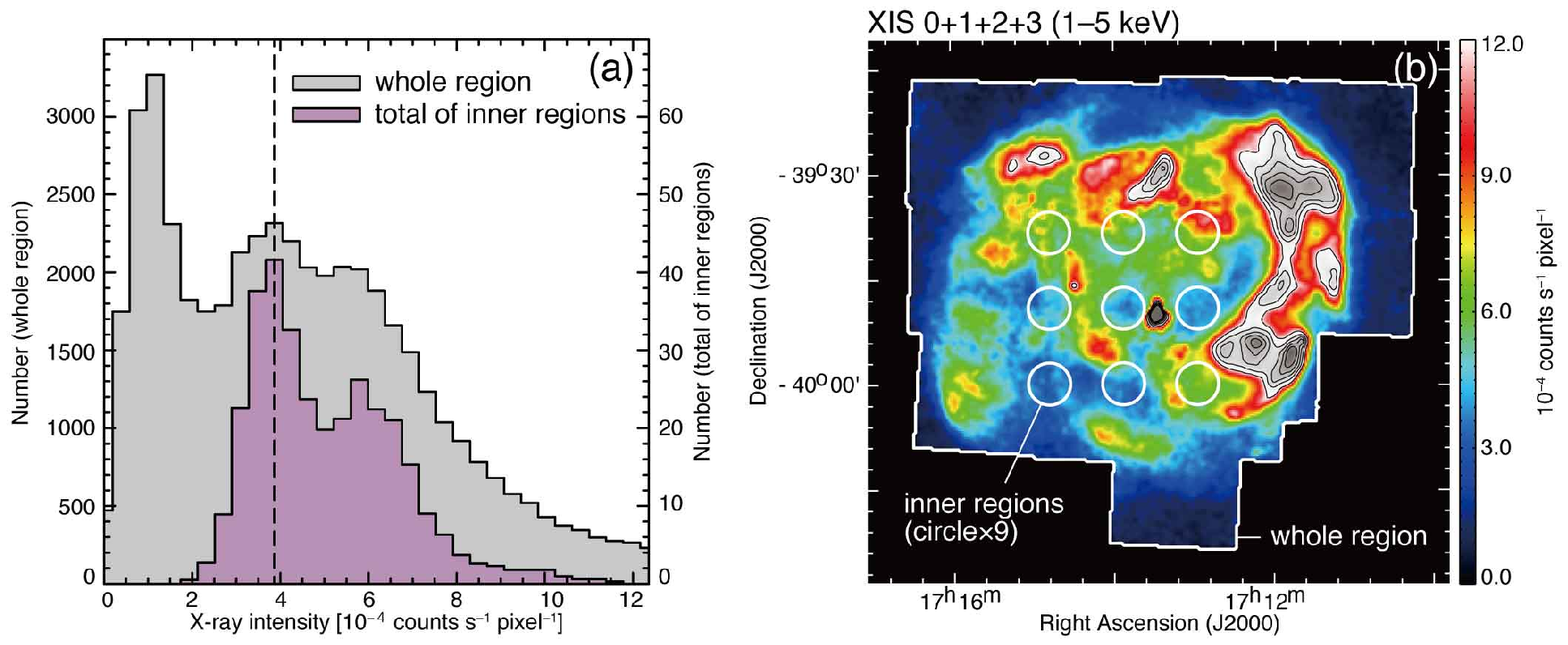}
\caption{(a) Histograms of X-rays intensity in the energy band 1--5 keV. The gray and magenta histograms show the contributions of the whole observed region and of the typical inner region of the SNR, respectively. (b) Same XIS mosaic image (1--5 keV) as Figure \ref{fig2} (c). The region enclosed by the broken solid lines and that enclosed by the nine circles represent the whole and typical inner region used for (a), respectively.}
\label{fig13}
\end{center}
\end{figure*}%

\end{document}